\documentclass[a4paper,fleqn,usenatbib]{mnras}

\usepackage{mathptmx}
\usepackage[T1]{fontenc}
\usepackage{ae,aecompl}

\usepackage{graphicx}	% Including figure files
\usepackage{amsmath}	% Advanced maths commands
\usepackage{amssymb}	% Extra maths symbols
\usepackage{multirow}

\newcommand{\cmnt}[1]{}
\newcommand{\review}[1]{#1}

\newcommand{\software}[1]{\texttt{\sc #1}}

\newcommand{\porb}{$P$}
\newcommand{\pdot}{$\dot{P}$}
\newcommand{\mdot}{$\dot{M}$}

\newcommand{\filus}{\textit{u$_s$}}	
\newcommand{\filgs}{\textit{g$_s$}}	
\newcommand{\filrs}{\textit{r$_s$}}	
\newcommand{\filis}{\textit{i$_s$}}	
\newcommand{\filzs}{\textit{z$_s$}}	
\newcommand{\filu}{\textit{u'}}	
\newcommand{\filg}{\textit{g'}}	
\newcommand{\filr}{\textit{r'}}	
\newcommand{\fili}{\textit{i'}}	
	
\newcommand{\filkg}{\textit{KG5}}	

\newcommand{\ou}{$^1$}
\newcommand{\cu}{$^2$}
\newcommand{\warwick}{$^3$}
\newcommand{\vienna}{$^4$}
\newcommand{\uva}{$^5$}
\newcommand{\ctac}{$^6$}
\newcommand{\yunnan}{$^7$}
\newcommand{\ioa}{$^8$}
\newcommand{\hamburg}{$^{9}$}
\newcommand{\ljmu}{$^{10}$}
\newcommand{\narit}{$^{11}$}
\newcommand{\sheffield}{$^{12}$}
\newcommand{\iac}{$^{13}$}
\newcommand{\sheffieldrse}{$^{14}$}
\newcommand{\ucc}{$^{15}$}

\title[No Period Change in Two AM\,CVn Binaries]{No Period Change in Two Long-Period AM\,CVn Binaries}

\author[M. J. Green]{
Matthew J. Green\ou$^,$\cu\thanks{E-mail: mjgreenastro@gmail.com},
Thomas R. Marsh\warwick,
Jan van Roestel\vienna$^,$\uva,
\newauthor
Tin Long Sunny Wong\ctac,
Diogo Belloni\yunnan,
Mukremin Kilic\ou,
Elm\'{e} Breedt\ioa,
\newauthor
Alex Brown\hamburg,
Chris M. Copperwheat\ljmu,
Anurak Chakpor\narit,
V. S. Dhillon\sheffield$^,$\iac,
\newauthor
Noel Castro Segura\warwick,
Martin J. Dyer\sheffield$^,$\sheffieldrse,
James Garbutt\sheffield,
Dan Jarvis\sheffield,
\newauthor
Vasu Kengkriangkrai\narit,
Mark R. Kennedy\ucc,
Paul Kerry\sheffield,
Thomas Kupfer\hamburg,
\newauthor
S. P. Littlefair\sheffield,
James McCormac\warwick,
James Munday\warwick,
Steven G. Parsons\sheffield,
\newauthor
Eleanor Pike\sheffield,
Ingrid Pelisoli\warwick,
Pablo Rodr\'{i}guez-Gil\iac,
David I. Sahman\sheffield,
\newauthor
and Amalie Yates\sheffield.
\\
\ou Homer L. Dodge Department of Physics and Astronomy, University of Oklahoma, 440 W. Brooks Street, Norman, OK 73019, USA
\\
\cu JILA, University of Colorado and National Institute of Standards and Technology, 440 UCB, Boulder, CO 80309-0440, USA
\\
\warwick Astronomy and Astrophysics Group, Department of Physics, University of Warwick, Coventry, CV4 7AL, United Kingdom
\\
\vienna Institute of Science and Technology Austria, Am Campus 1, Klosterneuburg, 3400, Austria
\\
\uva Anton Pannekoek Institute for Astronomy, University of Amsterdam, 1090 GE Amsterdam, The Netherlands
\\
\ctac The Observatories of the Carnegie Institution for Science, Pasadena, CA 91101, USA
\\
\yunnan International Centre of Supernovae (ICESUN), Yunnan Key Laboratory of Supernova Research, Yunnan Observatories, CAS,\\\quad Kunming 650216, China
\\
\ioa Institute of Astronomy, University of Cambridge, Madingley Road, Cambridge, CB3~0HA, United Kingdom
\\
\hamburg Hamburger Sternwarte, University of Hamburg, Gojenbergsweg 112, 21029 Hamburg, Germany
\\
\ljmu Astrophysics Research Institute, Liverpool John Moores University, IC2, Liverpool Science Park, 146 Brownlow Hill, Liverpool\\\quad L3 5RF, United Kingdom
\\
\narit National Astronomical Research Institute of Thailand (Public Organization), 260 Moo 4, Donkaew, Mae Rim, Chiang Mai, 50180,\\\quad Thailand
\\
\sheffield Astrophysics Research Cluster, School of Mathematical and Physical Sciences, University of Sheffield, Sheffield, S3~7RH,\\\quad United Kingdom
\\
\iac Instituto de Astrof\'isica de Canarias, 38205 La Laguna, Tenerife, Spain
\\
\sheffieldrse Research Software Engineering, University of Sheffield, Sheffield, S1~4DP, United Kingdom
\\
\ucc School of Physics, University College Cork, Cork, T12 K8AF, Ireland
}

\date{Accepted XXX. Received YYY; in original form ZZZ}

\pubyear{2020}

\begin{document}
\label{firstpage}
\pagerange{\pageref{firstpage}--\pageref{lastpage}}
\maketitle

\begin{abstract}

Ultracompact binary systems, consisting of two compact objects in an orbit $\lesssim 0.5 R_\odot$, should exhibit measurable rates of orbital period change ($\dot{P} \neq 0$) due to the emission of gravitational waves (GWs).
Measurements of \pdot\ have so far been limited to the shortest-period ultracompact binaries ($\lesssim 20$\,min).
Among the AM\,CVn-type subclass, several works have proposed the presence of extra angular momentum loss beyond GW emission, with magnetic braking being a widely discussed mechanism.
If present, this magnetic braking would dominate the angular momentum loss of AM\,CVn-type binaries with orbital periods $\gtrsim 30$\,min.
In this work, we present a long-term eclipse timing study of two AM\,CVn-type binaries, YZ\,LMi and Gaia14aae, with respective orbital periods of 28.3\,min and 49.7\,min and continuous observations since 2006 and 2015.
Both systems show $\dot{P}$ consistent with zero within $2\sigma$. 
Their $3\sigma$ upper limits are $1.1 \times 10^{-13}\,{\rm s \, s}^{-1}$ and $9.7 \times 10^{-14}\,{\rm s \, s}^{-1}$ respectively.
These non-detections are most simply explained by a scenario in which secular angular momentum loss is not substantially stronger than GW emission at all orbital periods, but is combined with deviations from the secular $\dot{P}$ whose timescales span decades but whose amplitude is $\lesssim 10^{-13}\,{\rm s \, s}^{-1}$. %, orders of magnitude smaller than the eclipse timing variations seen in hydrogen-dominated cataclysmic variables.
Our non-detections of $\dot{P}$ represent a limit on the strength of any enhanced angular momentum loss beyond pure GW emission.

% word count 243 -- Jan 9 2026

\end{abstract}


%%%%%%%%%%%%%%%%%%%%%%%%%%%%%%%%%%%%%%%%%%%%%%%%%%

%%%%%%%%%%%%%%%%% BODY OF PAPER %%%%%%%%%%%%%%%%%%

\section{Introduction}

\begin{table*}
\caption{AM\,CVn binaries with measured \pdot\ in the literature. We also show the mass ratio $q$ for systems for which it has been measured, and derive the donor response to mass loss $\xi$ according to the discussion in Section~\ref{sec:discussion:period-bounce}.
We note that the mass ratio of HM\,Cnc is unclear; we quote here values from both \citet{Roelofs2010} and \citet{Munday2023}, and the implied $\xi$ in each case.
We also note that \citet{Chakraborty2024} and \citet{Chickles2026} find lower limits on the donor and accretor masses for their three systems, but these do not necessarily translate to a constraint on $q$ and therefore $\xi$.
}
\begin{tabular}{lccccc}
\label{tab:literature}
Target & \porb\ & \pdot\ & $q$ & $\xi$ & References \\
 & [min] & [$\times 10^{-12}$\,s\,s$^{-1}$] & \\
\hline
\multirow{2}*{HM\,Cnc} & \multirow{2}*{5.35} & \multirow{2}*{$-36.57 \pm 0.01$} & $0.50 \pm 0.13^*$ & $1.1 \pm 0.4$ & \citet{Israel1999}, \citet{Roelofs2010}, \\
&&& $\approx 0.15$--$0.21^*$ & $\gtrsim5$ & \citet{Kaplan2012}, \citet{Munday2023} \\
ZTF\,J0546+3843 & 7.95 & $-4.3 \pm 1.1$ & -- & -- & \citet{Chakraborty2024} \\
ATLAS\,J1013$-$4516 & 8.56 & $-1.60 \pm 0.07$ & -- & -- & \citet{Chickles2026} \\
ZTF\,J1858$-$2024 & 8.68 & $7.8 \pm 1.4$ & -- & -- & \citet{Chakraborty2024} \\
V407\,Vul & 9.48 & $-2.27 \pm 0.25$ & -- & -- & \citet{Motch1996}, \citet{Strohmayer2004} \\
ES\,Cet & 10.3 & $3.2 \pm 0.1$ & -- & -- & \citet{Warner2002}, \citet{deMiguel2018} \\
% ZTF\,J0127+5258 & 13.7 & $-6.53 \pm 0.19$ & -- & -- & \citet{Burdge2023}\\
AM\,CVn & 17.1 & $0.85 \pm 0.48$ & $0.18 \pm 0.01$ & $0.08 \pm 0.11$ & \citet{Smak1967,Smak2023} \\
\hline
\end{tabular}
\end{table*}

\begin{table*}
\caption{Targets observed in this work. Accretor and donor masses are represented by $M_1$ and $M_2$ respectively.}
\begin{tabular}{lccccc}
\label{tab:targets}
Target & Coordinates & \porb\ & $M_1$ & $M_2$ & References \\
 &(J2000)& [min] & [$M_\odot$] & [$M_\odot$] \\
\hline
YZ\,LMi & 09:26:38.72~+36:24:02.5 & 28.3 & $0.85 \pm 0.04$ & $0.035 \pm 0.003$ & \citet{Anderson2005}, \citet{Copperwheat2011} \\
Gaia14aae & 16:11:33.97~+63:08:31.9 & 49.7  & $0.87 \pm 0.02$ & $0.025 \pm 0.0013$ & \citet{Campbell2015}, \citet{Green2018a} \\
\hline
\end{tabular}
\end{table*}

The emission of gravitational waves (GWs) drives the orbital period evolution of binary systems in very compact orbits ($\lesssim 0.5 R_\odot$).
The GW-driven mergers of extragalactic black hole and neutron star binaries are now regularly detected by ground-based detectors \citep[e.g.][]{Abbott2016,Abbott2023}.
Galactic compact binaries will be detected by future space-based GW observatories such as \textit{LISA} \citep[the Laser Interferometer Space Antenna;][]{Amaro-Seoane2017} or \textit{TianQin} \citep{Luo2016}.
Individual binary systems that \textit{LISA} will resolve can be used as calibration sources, while other, unresolved binaries will form a noise floor for \textit{LISA} observations \citep[e.g.][]{Korol2022,Kupfer2024}.

The dominant Galactic GW sources for these detectors will be ``ultracompact binary systems'' (those with orbital periods $P \lesssim$ 1\,h).
The majority of these belong to one of two categories: non-accreting double white dwarfs (DWDs), and a class of mass-transferring binary known as AM\,CVn-type binaries. 
In the latter class, each system consists of a white dwarf accreting matter from an evolved donor star which may be a low-mass white dwarf, semi-degenerate helium star, or the stripped core of a partially-evolved cataclysmic variable (CV) donor \citep[see][for a recent overview]{Green2025}. 
Open questions surround the evolutionary history of AM\,CVn-type binaries, and are closely related to the uncertain nature of their donor stars.

Although GWs from Galactic binaries have not yet been directly detected, the GW-driven change in orbital period has been detected for a number of ultracompact binaries.
Among detached DWDs, a number of known systems with $P \lesssim 20$\,min have a measured rate of period change which is constant within measurement uncertainties \citep[e.g.][]{Hermes2012,Burdge2019,Burdge2019b,Burdge2020}.
For those systems which also have independent measurements of the component masses, the rate of orbital period change (\pdot) 
is  consistent within uncertainties with the GW prediction \citep{Hermes2012,Burdge2019b}. 
A discrepancy of at least a few per cent from tidal dissipation is expected, but not yet measurable \citep{Burdge2019,Burdge2019b,Piro2019}.

% differs from the GW prediction by only a few per cent, a discrepancy that is of the order expected from tidal dissipation \citep[e.g.][]{Piro2011,Piro2019}.

For mass-transferring binaries, the picture is more complicated. 
In this case, \pdot\ can be understood as the result of two opposing processes: angular momentum loss (AML), which will tend to drive the system to shorter orbital periods, and the redistribution of mass from the donor to the accretor, which will tend to drive the system to longer orbital periods.
A number of AM\,CVn-type binaries with $P \lesssim 20$\,min have detections of \pdot, which we list in Table~\ref{tab:literature}.\footnote{
Not listed in Table~\ref{tab:literature} are two further mass-transferring binaries.
ZTF\,J0127+5258 has a warm donor with visible hydrogen, unlike any other AM\,CVn-type system, and is inspiralling with $\dot{P} = (-6.5 \pm 0.2) \times 10^{-12}\,{\rm s \, s}^{-1}$ \citep{Burdge2023}. 
ZTF\,J1813+4251 is from the class of binaries sometimes called ``hybrid systems'', which have traits in common with both AM\,CVn-type systems and hydrogen-rich CVs, and has a non-detection of $\dot{P} = (-0.3 \pm 1.2) \times 10^{-12}\,{\rm s \, s}^{-1}$ \citep[][]{Burdge2022}.
The \pdot\ of both systems are interesting, but their nature appears to be somewhat different from the AM\,CVn systems discussed in this work and so we do not include them in our analysis.
}
% All of these \pdot\ values are also listed in the catalogue of ultracompact binary systems maintained by \citet{Green2025}.
There is significant scatter among these, notably including both positive values (corresponding to increasing orbital periods) and negative values (corresponding to decreasing orbital periods).
In the case of the shortest-period system, HM\,Cnc, a second period derivative $\ddot{P}$ has also been measured \citep[][]{Strohmayer2021,Munday2023}.

The expected \pdot\ behaviour of an AM\,CVn-type binary depends on both the assumed formation scenario and what assumptions are made about the AML from the binary.
Formation scenarios have been traditionally separated into three groups, named for their immediate progenitors: the white dwarf donor channel, in which the progenitor is a detached DWD; the helium star donor channel, in which the progenitor is a detached binary consisting of a white dwarf and a hot subdwarf; and the evolved CV channel, in which the progenitor is a CV with an evolved (typically subgiant) donor.

In scenarios with white dwarf or helium star donors, mass transfer begins at short periods ($P \lesssim 15$\,min) and the binary then evolves to longer periods \citep{Paczynski1967,Savonije1986,Iben1987,Deloye2007,Yungelson2008,Wong2021,Bauer2021,Chen2022,Rajamuthukumar2025}.
In general, the magnitude of \pdot\ under these scenarios should be large among newly formed AM\,CVn systems with short orbital periods, and smaller for systems with longer orbital periods.
For a degenerate donor that expands in response to mass loss, one would na\"{i}vely expect \pdot\ to be strictly positive.
In order to explain the observed negative \pdot\ values, models have invoked a short-lived turn-on phase in which a non-degenerate outer atmosphere of the donor star is being removed, and mass transfer rate is increasing but is not yet sufficiently large to dominate the \pdot\ \citep{DAntona2006,Deloye2007,Kaplan2012}.
It is unclear what fraction of short-period AM\,CVn binaries should have negative \pdot\ under this model.

In the evolved CV scenario, the binary begins stable mass transfer from a subgiant donor onto a white dwarf \citep[e.g.][]{Podsiadlowski2003,Goliasch2015,Belloni2023,Sarkar2023a,Sarkar2023}.
It evolves inward from longer orbital periods, and (given the appropriate initial conditions) passes through a period bounce and thereafter evolves outwards.
The position of this period bounce and subsequent \pdot\ behaviour depends in part on the AML model, which
is often assumed to be dominated by GW radiation. However, some models \citep[e.g.][]{Belloni2023,Sarkar2023a,Sarkar2023} invoke additional AML through magnetic braking that dominates over the GW radiation through some phases of evolution.
These models have attracted significant attention in recent years, as they readily explain observational results such as large donor radii \citep[e.g.][]{Green2018a,Green2018,vanRoestel2022} that are a challenge for other models.
Under evolved CV models in which AML is dominated by GW radiation, the period bounce typically occurs at $40 \lesssim P \lesssim 70$\,min \citep[e.g.][]{Goliasch2015}, but additional magnetic braking may reduce this to $10 \lesssim P \lesssim 50$\,min \citep[e.g.][]{Belloni2023}.
In either case, the period bounce approximately coincides with or precedes the depletion of hydrogen below detectable levels in the accreted material.
Given that most mass-transferring ultracompact binaries do not have detectable levels of hydrogen, it may be expected that most are either near the period bounce or have \pdot\,$> 0$ under this formation scenario.
In models with AML by pure GW radiation, \pdot\ should be maximal shortly after the period bounce and monotonically decrease with increasing orbital period, but in models which invoke extra AML by magnetic braking, larger values of \pdot\ may be expected at longer orbital periods.

To date, all \pdot\ measurements in AM\,CVn-type binaries are concentrated at the shortest end of the orbital period distribution, where the GW radiation is greatest.
Further information can be gained by extending \pdot\ measurements to longer-period systems. In this context, eclipsing binaries provide ideal testing grounds for predictions of \pdot\ for two reasons. 
Firstly, the eclipse gives a sharp photometric signature for the orbital period, allowing for more precise timing measurements than can be achieved in non-eclipsing binaries.
Secondly, the eclipse also makes it possible to derive precise measurements of the component stellar masses (under an assumed mass-radius relationship for the accretor), and hence predict the AML due to GW emission.

Over ten long-period ($P>20$\,min) eclipsing AM\,CVn binaries are currently known \citep[e.g.][]{vanRoestel2022,Rodriguez2023,Khalil2024},
but only two have been observed over a sufficient timespan to make a \pdot\ measurement feasible.
YZ\,LMi (also known as SDSS\,J0926+3624) has an orbital period of 28.3\,min and has been known since 2005 \citep{Anderson2005, Copperwheat2011}.
Gaia14aae (also known as ASASSN-14cn) has an orbital period of 49.7\,min and was discovered in 2014 \citep{Campbell2015,Green2018a,Green2019}.
Both systems have mass measurements available in the literature and have been observed continuously since their discovery.
Key details of these binaries are listed in Table~\ref{tab:targets}.
A previous measurement of \pdot\ has been suggested in the literature for YZ\,LMi, based on three epochs of observations between 2006 and 2012 \citep{Szypryt2014}; however, as we will show in the following sections, more recent observations are inconsistent with that measurement.

In this paper, we present eclipse timing measurements of YZ\,LMi and Gaia14aae, covering the timespans 2006--2024 for YZ\,LMi and 2015--2024 for Gaia14aae.
These measurements are derived from observations with ULTRACAM, ULTRASPEC, HiPERCAM, and CHIMERA \citep{ultracam,ultraspec,hipercam,chimera}.
Section~\ref{sec:obs} describes the observational set-up used for all observations. 
Section~\ref{sec:timing} describes the process used to measure the time of each eclipse, and Section~\ref{sec:ephemeris} presents the mid-eclipse times and the derived ephemerides.
Section~\ref{sec:predictions} then compares the measured eclipse times to theoretical predictions, and discusses possible sources of the disagreement between the two.

\section{Observations}
\label{sec:obs}

High-speed photometric observations of YZ\,LMi and Gaia14aae were collected over periods of 19 and 10 years, respectively. 
The observations of YZ\,LMi are listed in Tables~\ref{tab:observations-yzlmi-1}--\ref{tab:observations-yzlmi-2} and the observations of Gaia14aae are shown in Table~\ref{tab:observations-gaia14aae}.
Some of these data have been previously published: observations of YZ\,LMi up to 2009 were published by \citet{Copperwheat2011} and observations of Gaia14aae up to 2017 were published by \citet{Green2018a}.
Several observations of YZ\,LMi coincided with dwarf nova outbursts; these are marked in Tables~\ref{tab:observations-yzlmi-1}--\ref{tab:observations-yzlmi-2} and discussed in later sections.

These data were collected using four high-speed cameras: ULTRACAM \citep{ultracam}, ULTRASPEC \citep{ultraspec}, CHIMERA \citep{chimera}, and HiPERCAM \citep{hipercam}.
All four are imaging photometers which utilise frame-transfer CCDs to minimise dead time between exposures. 
ULTRACAM collects data simultaneously in three different colour bands, while ULTRASPEC is single-band, CHIMERA is dual-band, and HiPERCAM has five bands. 

For these observations ULTRACAM was mounted on the 4.2\,m William Herschel Telescope (WHT) at the Observatorio del Roque de los Muchachos on La Palma, Spain. 
ULTRASPEC was mounted on the 2.4\,m Thai National Telescope (TNT) on Doi Inthanon, Thailand.
CHIMERA was mounted on the 5.1\,m Hale Telescope at Palomar Observatory, USA.
HiPERCAM was mounted on the 10.4\,m Gran Telescopio Canarias (GTC), also on La Palma.

For our observations with ULTRACAM and CHIMERA, the filters from the standard Sloan \filu\filg\filr\fili\ filter set were used.
With ULTRASPEC, we instead used the custom \filkg\ filter, a broad, blue filter that was chosen to maximise through-put \citep[][appendix]{Hardy2017}.
For HiPERCAM, the custom ``Super-SDSS'' \filus\filgs\filrs\filis\filzs\ filters were used. These filters are designed to cover the same wavelength ranges as the standard Sloan filters, but with greater throughputs \citep{hipercam,Brown2022}.

Reductions from all four instruments were carried out with the standard ULTRACAM and HiPERCAM\footnote{\software{https://github.com/hipercam/hipercam}} pipelines.
Each image was bias-subtracted and divided throughout by a flat field in the same filter. 
Flat fields were used from the same night if available, or otherwise the closest available flat field was used.
Fluxes were extracted from each frame using aperture photometry, with a variable-width aperture set to a radius of 1.8 $\times$ the measured full-width half-maximum (FWHM) of stars in that image.
In each image, the flux of the target was divided by the flux of a nearby, non-variable comparison star to correct for atmospheric effects.
%Fluxes were calibrated against the same nearby stars.
The comparison stars used for YZ\,LMi were 09:26:54.0 +36:23:52, 09:26:34.2 +36:25:13, and 09:26:39 +36:23:35. 
For Gaia14aae they were 16:11:30.5 +63:09:26, 16:11:29.0 +63:08:08, 16:11:21 +63:08:02, and 16:11:08 +63:09:40.

%
%Outbursts affecting our data on:
%2012-01-14--20
%2017-12-05--12
%2023-03-14--15

\section{Eclipse timing measurements}
\label{sec:timing}

\begin{table}
\caption{Mid-eclipse times of YZ\,LMi. The first lines are shown here for demonstrative purposes; the full table is available in the online material.}
\begin{tabular}{ccc}
\label{tab:measurements-yzlmi}
Filter & Mid-eclipse time & Uncertainty \\
& [BMJD(TDB)] & \\
\hline 
$u'$ & 53795.945516 & 0.000092 \\
$g'$ & 53795.945497 & 0.000085 \\
$r'$ & 53795.945491 & 0.000095 \\
$u'$ & 53795.965177 & 0.000101 \\
$u'$ & 53795.984832 & 0.000107 \\
$g'$ & 53795.965156 & 0.000070 \\
$g'$ & 53795.984832 & 0.000061 \\
$r'$ & 53795.965154 & 0.000078 \\
$r'$ & 53795.984829 & 0.000074 \\
$u'$ & 53796.004502 & 0.000069 \\
\vdots & \vdots & \vdots\\
\hline
\end{tabular}
\end{table}

\begin{table}
\caption{As Table~\ref{tab:measurements-yzlmi}, but for Gaia14aae.}
\begin{tabular}{ccc}
\label{tab:measurements-gaia14aae}
Filter & Mid-eclipse time & Uncertainty \\
& [BMJD(TDB)] & \\
\hline 
$u'$ & 57037.220079 & 0.000010 \\
$u'$ & 57037.254592 & 0.000008 \\
$u'$ & 57037.289104 & 0.000007 \\
$g'$ & 57037.220072 & 0.000006 \\
$g'$ & 57037.254589 & 0.000004 \\
$g'$ & 57037.289104 & 0.000003 \\
$i'$ & 57037.220074 & 0.000010 \\
$i'$ & 57037.254596 & 0.000011 \\
$i'$ & 57037.289114 & 0.000008 \\
$u'$ & 57038.186609 & 0.000007 \\
\vdots & \vdots & \vdots\\
\hline
\end{tabular}
\end{table}

%\subsection{Eclipse timing measurements}
\begin{figure*}
\includegraphics[width=500pt]{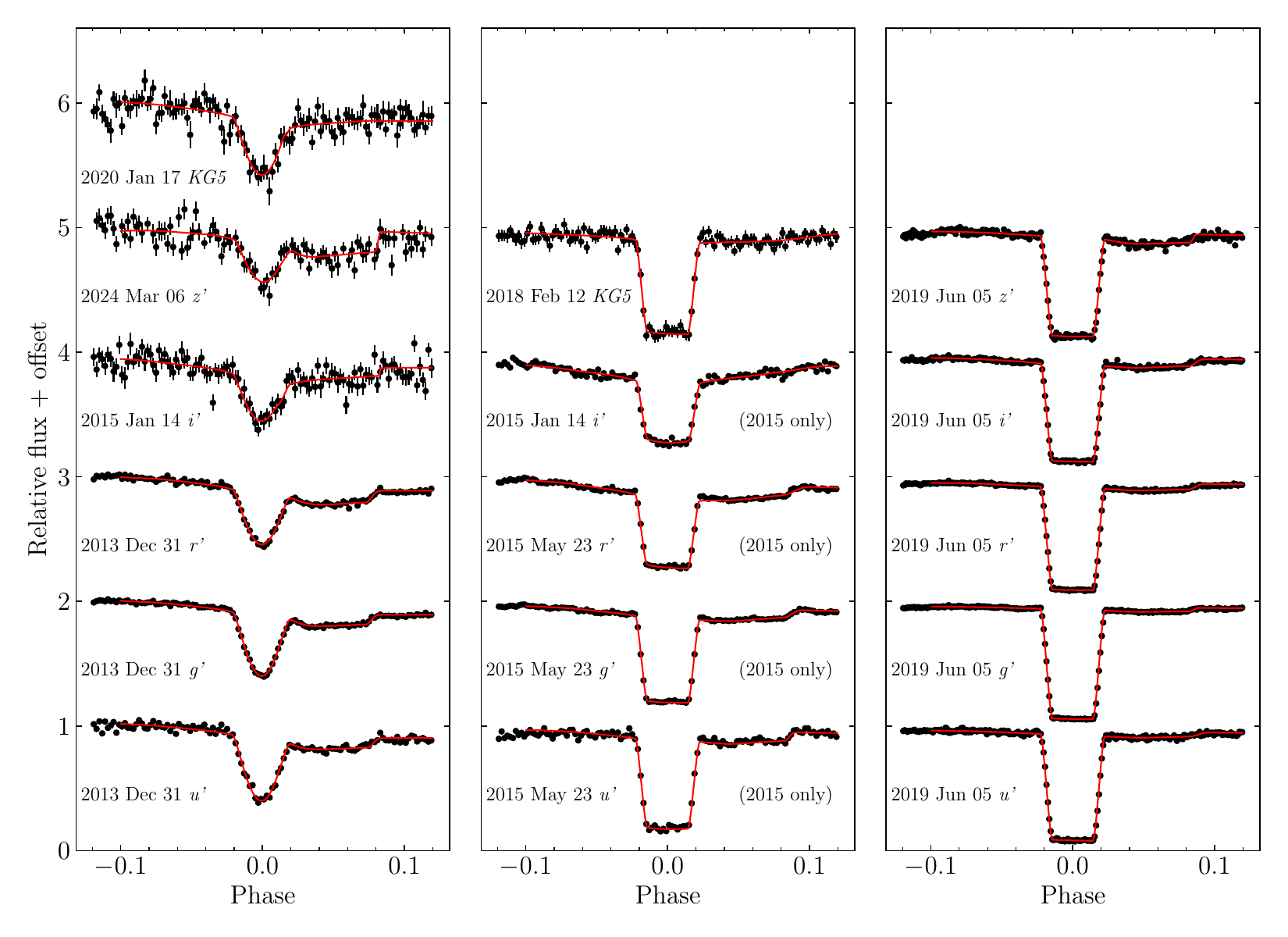}
\caption{Template eclipse profiles for YZ\,LMi (left) and Gaia14aae (centre and right), each plotted against one night of phase-folded and binned data. The four profiles marked `2015 only' were derived to account for the shallower eclipses of Gaia14aae in 2015, as discussed in the text.
}
\label{fig:lcurves}
\end{figure*}

\begin{figure}
\includegraphics[width=\columnwidth]{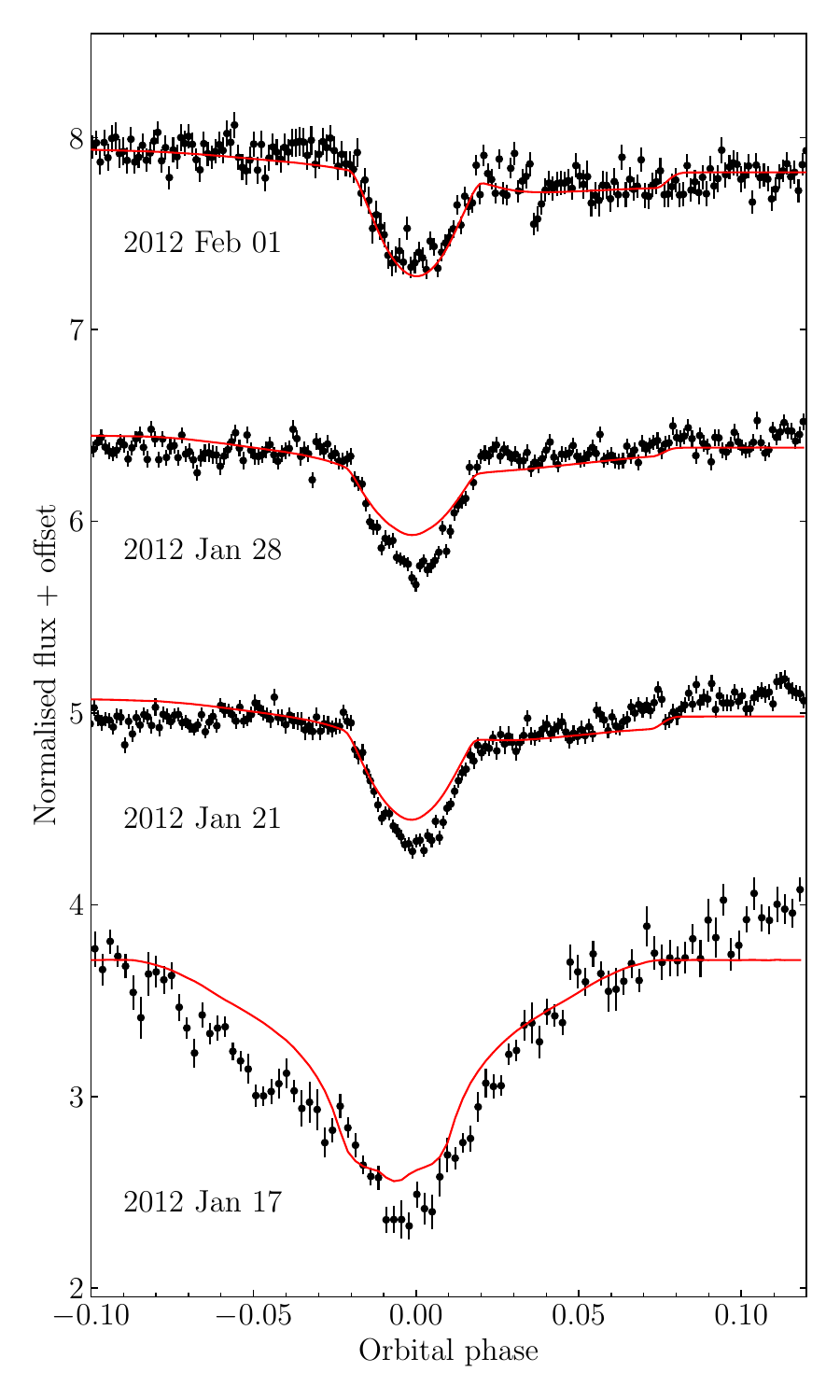}
\caption{Example eclipse light curves of YZ\,LMi from several nights in 2012 Jan--Feb, following a dwarf nova outburst in mid-Jan. 
Although the out-of-eclipse flux had returned to normal by Jan 21, the eclipse profile remained substantially altered until Feb 01.
Mid-eclipse times measured prior to Feb 01 were noticeably shifted from a linear ephemeris due to this unusual eclipse profile. 
We therefore discarded all mid-eclipse times measured from 2012 Jan.
}
\label{fig:lc-outburst}
\end{figure}

In order to measure the mid-eclipse time ($t_0$) of each eclipse, individual individual eclipses were separately fitted in each passband using a model lightcurve for which all parameters were held fixed other than $t_0$ (and component fluxes in the case of Gaia14aae, as discussed below).
The eclipse lightcurves were modelled using the \software{lcurve} software package \citep{Copperwheat2010}.
This program models four luminous components of the binary system: the accreting white dwarf, the accretion disc, the bright spot at the point of intersection between the accretion stream and the disc, and the Roche lobe-filling donor star. In our case the contribution from the donor star was assumed to be negligible.
Parameters in the model include the mid-eclipse time, the orbital period, the component masses and temperatures of the two stars, the radius of the accretor, the orbital inclination of the system relative to the line of sight from the Earth, and a set of parameters to describe the accretion disc (including its radius and temperature profile) and the bright spot (including its temperature, scale length, and distance from the primary star). 
Note that the `temperature' parameters in this model are in effect proxies for the flux of a component in a given bandpass, and are only rough approximations for the physical temperatures. 
More detailed descriptions of the model can be found in \citet{Copperwheat2010} and \citet{Green2018a}.

We first sought a set of models that would describe the eclipse profile in each bandpass. 
A phase-folded light curve was produced in each bandpass, as shown in Fig.~\ref{fig:lcurves}.
For the purposes of this study, the physical meanings of model parameters other than the mid-eclipse time are not of interest; we only require that the model describes the eclipse profile well.
We therefore held fixed all physical parameters of the component stars and orbit (namely orbital period and inclination, stellar masses, radii, and temperatures) at their published values from \citet{Copperwheat2011} and \citet{Green2018a}.
Parameters describing the temperature of the accretion disc and the shape and size of the bright spot were allowed to vary so that a good fit to the shape of those eclipse profiles was obtained.
Models were then converged on the phase-folded data in each bandpass using a  combination of simplex and Levenberg-Marquardt algorithms.\footnote{If the physical parameters of the system are of interest it is advised to use a more robust fitting procedure such as the  Markov Chain Monte Carlo procedure used by \citet{Copperwheat2011} and \citet{Green2018a}.
}
% These converged models were then used as the starting point for more constrained fits to individual eclipses, as described later in this section.

These filter-dependent models discussed above were then used to measure $t_0$ for each individual eclipse. 
Separate light curves were produced for each eclipse in each bandpass.
A Levenberg-Marquardt algorithm was used to fit the model to the data, holding all parameters fixed except for $t_0$ (and the white dwarf flux in the case of Gaia14aae).
Each fit was inspected by eye to ensure that it was of a reasonable quality, and eclipses where the fit was clearly poor were thrown out. 
Poor quality fits were due to several causes: low signal-to-noise (resulting from poor observing conditions, most often clouds); the intrinsic red noise of the system known as `flickering' which can sometimes substantially distort the profile of the eclipse; dwarf nova outbursts (see below); and eclipses that were only partially covered by observations.
% The 2009 eclipses during which YZ\,LMi was in outburst were also discarded.
Across all eclipse-passband combinations that passed this visual inspection, we were left with 447 mid-eclipse time measurements for YZ\,LMi and 181 for Gaia14aae.
These mid-eclipse times are summarised in Tables~\ref{tab:measurements-yzlmi} and \ref{tab:measurements-gaia14aae}.

Several observations of YZ\,LMi coincided with dwarf nova outbursts, which adversely affect the fitting procedure.
Affected data from 2017 December 12 and 2023 March 14--15 were discarded.
In 2012 January--February, a long series of observations followed YZ\,LMi through its transition from outburst to quiescence. 
Although the flux of the target had returned to quiescent levels by January 21, the {\sc lcurve} eclipse profiles remained a visibly poor fit until the end of January (Fig.~\ref{fig:lc-outburst}).
We experimented with allowing the temperatures of individual components (the accretor, disc, and bright spot) to vary, but still were not able to produce good quality fits to these data.
We therefore discarded all data from 2012 January, but continued to use data from 2012 February.

Gaia14aae has only one recorded outburst, in 2014. 
Following this outburst, there was a notable change in the eclipse profiles of Gaia14aae throughout the data from 2015 and 2016.
We previously noted in \citet[][their figure~11]{Green2018a} a decreasing trend in the flux of the central white dwarf from 2015 to 2016, likely explained by its cooling following a period of heating during outburst. 
The system appears to have since stabilised in its cooler 2016 state.
We modified our fitting procedure in two ways to handle this variable eclipse profile.
Firstly, we created separate starting \filu\filg\filr\fili\ models to describe the deeper eclipses from 2015 and the shallower eclipses from 2016 onwards.
Secondly, we allowed the white dwarf eclipse depth to vary relative to the eclipse depths of the accretion disc and bright spot during the fits to individual eclipses of Gaia14aae. 
These fits therefore each had two free parameters: the white dwarf temperature and $t_0$.
These measures allowed us to produce fits that visually matched the eclipse profile for each individual eclipse of Gaia14aae.

%
% Skipping these lines as I didn't include this analysis in the end, but leaving it here in case I change my mind
%
%\subsection{Eclipse depth measurements}
%
%For the discussion in Section~\ref{sec:predictions}, it will be useful to consider the flux contributions of individual components of YZ\,LMi and whether they vary over time.
%For this purpose, we created phase-folded datasets from the long observing runs in ??? years, using data from the \filg\ band on ULTRACAM in each case.
%Later HiPERCAM observations were not used, because the filter change from \filg\ to \filgs\ produced a notable change in eclipse depth.
%
%Eclipse profile models were then converged on these folded datasets.
%As in the fitting of folded data described previously, we held physical parameters fixed at their published values, while allowing the flux contribution of each component and the various parameters describing the shape and size of the bright spot to vary.
%The flux contribution of the accretor, accretion disc, and bright spot were then extracted from the converged models by taking the out-of-eclipse modelled flux between phases 0.2 and 0.5 (this also avoids the `orbital hump' of beamed light from the bright spot).
%Note that this procedure recovers the full flux of the accretor, not the depth of the partial eclipse.
%This procedure does not recover any formal uncertainty on the fluxes, but the hypothetical changes in flux that we are searching for are large.
%The results are discussed in Section~??.
%
%

\section{Ephemerides}
\label{sec:ephemeris}

\begin{figure*}
\includegraphics[width=500pt]{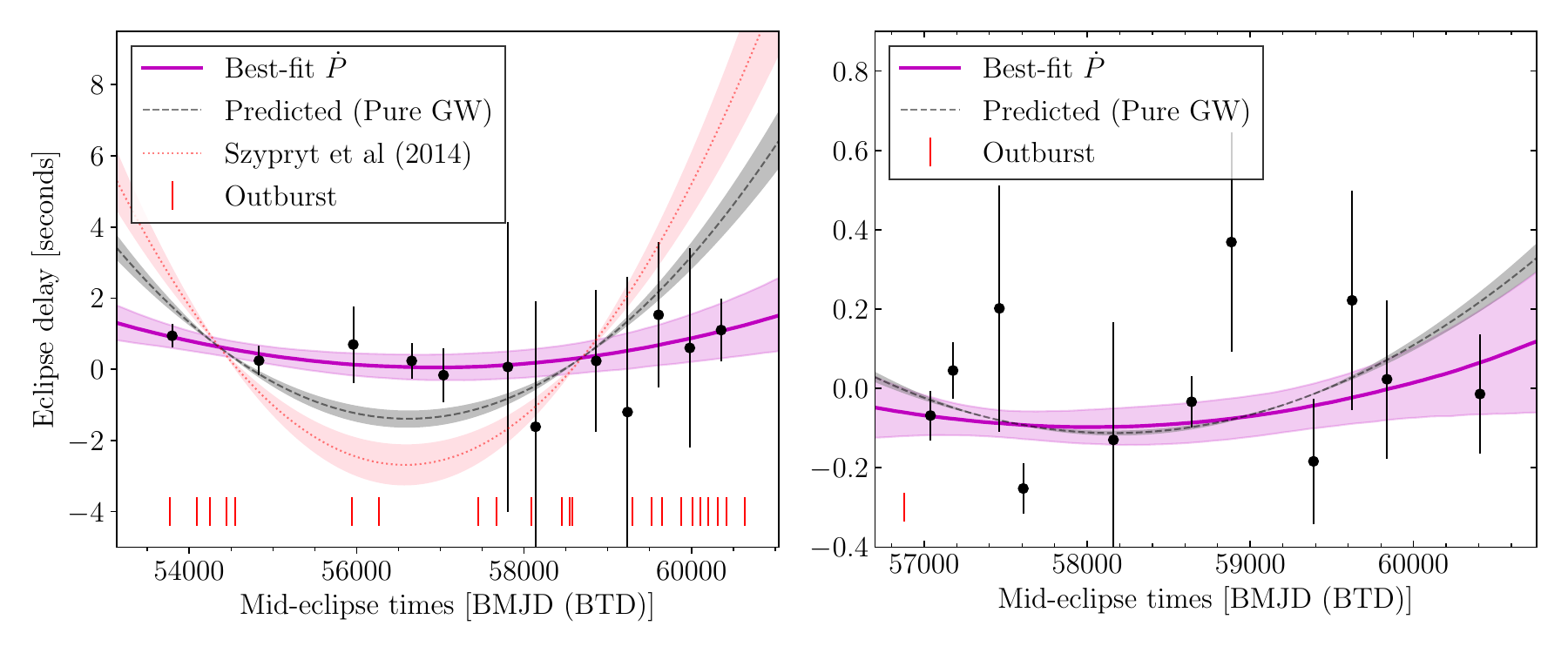}
\caption{Delay of the observed mid-eclipse times for YZ\,LMi (left) and Gaia14aae (right), compared to the expected mid-eclipse time under the assumption of a linear ephemeris.
For clarity the individual eclipse times have been binned in this figure, but the fit was performed to the unbinned data.
An unbinned version of this figure is shown in Fig.~\ref{fig:ephemeris-unbinned}.
Shaded regions show the 1\,$\sigma$ uncertainty region around each plotted line. 
% Dashed and dotted lines show predicted and previously measured period derivatives.
The times of known dwarf nova outbursts are indicated along the bottom of the figure.
For both systems, the best-fit \pdot\ values are consistent with a linear ephemeris within $2\sigma$.
}
\label{fig:ephemeris}
\end{figure*}

% \begin{figure}
% \includegraphics[width=\columnwidth]{figures/oc_zoom_2012.pdf}
% \caption{temp.
% }
% \label{fig:temp}
% \end{figure}

We first found the best-fit linear ephemeris for each binary system.
For YZ\,LMi, the best-fit linear ephemeris is
\begin{equation}
\mathrm{BMJD(TDB)} = 55469.3164759(6) + 0.019661273247(7) E,
\label{eq:eph-yzlmi}
\end{equation}
and for Gaia14aae it is
\begin{equation}
\mathrm{BMJD(TDB)} = 58073.8427825(3) + 0.03451957112(1) E,
\label{eq:eph-g14}
\end{equation}
where eclipses occur at integer values of the cycle number $E$.
Values in parentheses indicate the uncertainty in the final figure of the associated value.
For both of these ephemerides, the zero-point mid-eclipse time was chosen to minimise the correlation between mid-eclipse time and period.
These ephemerides represent minor corrections to the published orbital periods: by $30 \pm 2 \mu s$ for YZ\,LMi compared to the period of \citet{Copperwheat2011}, and $25 \pm 6 \mu s$  for Gaia14aae compared to the period of \citet{Green2018a}.

In order to test for the presence of a \pdot\ term in the measured mid-eclipse times, we fit a second-order polynomial to the residuals of the best-fit linear ephemeris for each of the two binary systems.
The fit was performed using an affine-invariant Markov Chain Monte Carlo (MCMC) algorithm implemented by the package {\sc emcee} \citep{Foreman-Mackey2013}.
The polynomial is interpreted as 
\begin{equation}
O - C = \frac{1}{2}\dot{P} \Delta P (E-E_0)^2 + \Delta P (E-E_0) + K
\end{equation}
where $O(E)$ is a given observed mid-eclipse time, $C(E)$ is the corresponding expected mid-eclipse time predicted by Equation~\ref{eq:eph-yzlmi} or \ref{eq:eph-g14}, $\Delta P$ is a correction to the linear orbital period, and $K$ and $E_0$ are vertical and horizontal offsets. 
$\Delta P$, $K$, and $E_0$ are all consistent with zero within $1 \sigma$ for both systems.%, but were included in the fit as they are possible in principle.

To account for any extra scatter that may be present in the mid-eclipse times but not included in the formal uncertainties ($\sigma_\mathrm{formal}$), we included an extra noise term ($\sigma_\mathrm{extra}$) such that the total uncertainty on each mid-eclipse time $i$ is $\sigma_i^2 = \sigma_{\mathrm{formal},i}^2 + \sigma_\mathrm{extra}^2 $. 
The extra noise term is subject to a prior whose probability is proportional to $1/\sigma_\mathrm{extra}$.
For YZ\,LMi the best-fit value of this extra noise term was $\sigma_\mathrm{extra} = 0.017$\,s, and for Gaia14aae it was $\sigma_\mathrm{extra} = 0.15$\,s.

Fig.~\ref{fig:ephemeris} shows the results of this MCMC fit.
For both binary systems, the \pdot\ term is within $2\sigma$ of zero.
For YZ\,LMi the best-fit value is $\dot{P} = (3.6 \pm 2.5) \times 10^{-14} {\rm s \, s}^{-1}$ or $1.2 \pm 0.8\,\mu \mathrm{s \, yr}^{-1}$.
For Gaia14aae it is $\dot{P} = (2.2 \pm 2.5) \times 10^{-14} {\rm s \, s}^{-1}$ or $0.7 \pm 0.8$\,$\mu \mathrm{s \, yr}^{-1}$. 
% Other terms in the polynomial fit are also consistent with zero.
We therefore conclude that there is no measurable change in period over the course of these observations.
The $3 \sigma$ upper limits on the period derivative are $1.1\times10^{-13} {\rm s \, s}^{-1}$ or $3.6 \, \mu \mathrm{s \, yr}^{-1}$ for YZ\,LMi and $9.8\times10^{-14} {\rm s \, s}^{-1}$ or $3.1 \, \mu \mathrm{s \, yr}^{-1}$ for Gaia14aae.

In the same figure we also show predicted \pdot\ curves based on a pure GW emission model (discussed in Section~\ref{sec:discussion:compare-models}), and the $\dot{P} = 9.7 \pm 1.8$\,$\mu s / yr$ measurement for YZ\,LMi made by \citet{Szypryt2014}. 
The latter measurement is in disagreement with our measurement at the $4\sigma$ level.
Their measurement was based on the 2006 and 2009 data from \citet{Copperwheat2011} which are also used here, combined with their own observations from 2012 December. 
While we cannot be certain, the broad eclipse profiles in the light curves in their figure 1 (particularly on the night of 2012 December 8) suggest that these observations coincided with an outburst.
In our own analysis of our data, before excluding the eclipses affected by the outburst of 2012 January, we found that those eclipses returned mid-eclipse times  later than expected by several seconds.
If a similar effect is present in the 2012 December eclipse time measurements of \citet{Szypryt2014}, this may explain the discrepancy between their result and ours.

\section{Comparison to Predictions}
\label{sec:predictions}

In this section we discuss the interpretation of our $\dot{P}$ measurements in the context of existing AM\,CVn formation models.
After an overview of the analytical background and some known issues in eclipse timing in other mass-transferring systems, we compare our measurement to AM\,CVn formation models with and without enhanced magnetic braking.
We rely on the model tracks from \citet{Wong2021} and \citet{Belloni2023}, both using MESA \citep[Modules for Experiments in Stellar Astrophysics;][]{Paxton2011}.
These works modeled separate formation scenarios, in which the progenitor system is a detached DWD or a CV with an evolved donor, respectively.
For the purposes of our discussion, the more relevant distinction is that \citet{Wong2021} assumed AML by pure GW radiation, while \citet{Belloni2023} included enhanced magnetic braking under the Convection And Rotation Boosted prescription \citep[CARB;][]{Van2019}.

% \cmnt{More details on M2 plot here}

\begin{figure}
\includegraphics[width=\columnwidth]{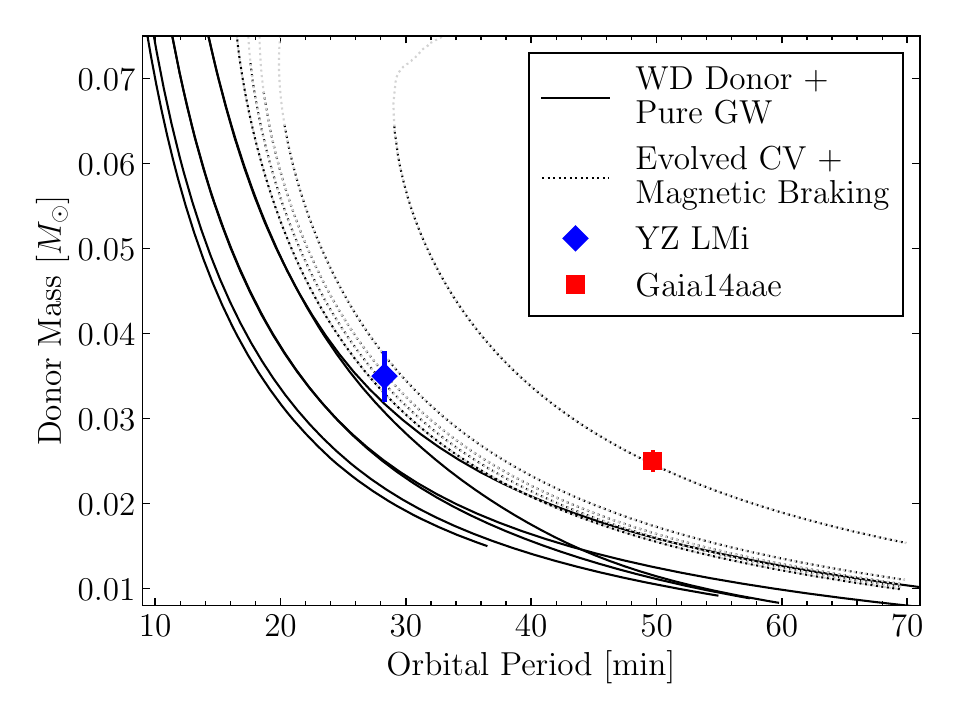}
\caption{Donor mass as a function of orbital period for our two targets, and for both sets of model tracks.
Pure-GW model tracks from \citet{Wong2021} struggle to reproduce the large donor mass observed for Gaia14aae, which is a known issue.
Model tracks with magnetic braking from \citet{Belloni2023} were chosen to reproduce the donor masses of our two targets.
Earlier sections of the \citet{Belloni2023} model tracks, in which the hydrogen content of the accreted material is greater than $10^{-3}$, are plotted in a fainter colour.
}
\label{fig:m2-p}
\end{figure}

The donor masses predicted by these model tracks are shown in Fig.~\ref{fig:m2-p}.
The models of \citet{Wong2021} include a range of initial donor entropies, leading to the vertical spread in tracks.
They also include two values for the efficiency of donor cooling.
Models with efficient donor cooling result in donor contraction (and hence lower $M_2$ for a given $P$) for $P \gtrsim 40$\,min.
However, even models with inefficient cooling, which retain larger $M_2$ at large $P$, do not successfully recreate the large $M_2$ measured for Gaia14aae.
These differences in initial values are not directly relevant for the discussion in this work; the tracks presented here are simply intended to cover the entire range of possible parameter space for $P$, $M_2$, and $\dot{P}$ (shown in later figures).

From \citet{Belloni2023}, we take five model tracks: four which reproduce the $M_2$ of YZ\,LMi and one which reproduces the $M_2$ of Gaia14aae.
For these model tracks, the vertical spread corresponds to the evolutionary state of the donor at the onset of mass transfer.
As discussed by \citet{Belloni2023}, the extra AML from magnetic braking is necessary in their models to reproduce the observed $M_2$ of Gaia14aae.
If not for their enhanced AML, the post-period bounce model tracks of \citet{Belloni2023} would evolve similarly to those of \citet{Wong2021} in the $\gtrsim 30$\,min orbital period range.

% For these purposes, it will be helpful to separate formation models of AM\,CVns into multiple categories. 
% The most relevant separation for our purposes is the difference between models in which AML is dominated by GW emission \citep[for which we will focus on the evolutionary tracks of][]{Wong2021}, and models in which AML is enhanced by substantial magnetic braking \citep[particularly the model tracks of][]{Belloni2023}.
% A separate distinction can be made between scenarios in which the direct progenitor to the AM\,CVn system is a detached double-degenerate binary (the white dwarf and helium star donor channels discussed in the Introduction) or in which stable Roche lobe overflow began with a non-degenerate donor (the evolved CV channel discussed in the Introduction).
% Enhanced magnetic braking may in principle be present under both scenarios, but it has particularly been invoked in the evolved CV scenario.

\subsection{Analytical background}
\label{sec:theory}

For the following section we refer to the extensive discussions of AM\,CVn evolution under the emission of gravitational waves by \citet{Marsh2004}, \citet{vanHaaften2012}, \citet{Sberna2021}, and \citet{Chakraborty2024}.
Any change in angular momentum stored in the orbit and spin of the donor ($\dot{J}_\mathrm{orbit}$, $\dot{J}_\mathrm{spin}$) can be understood as a combination of the AML from the binary through GW radiation ($\dot{J}_\mathrm{GW}$), the AML through magnetic braking ($\dot{J}_{\rm MB}$), the angular momentum of any material ejected from the system ($\dot{J}_\mathrm{eject}$), the angular momentum transferred from the donor star to the accretor via the accretion stream ($\dot{J}_\mathrm{stream}$), and the transfer of angular momentum from the accretor to the orbit via a tidal torque between the accretor and the donor ($\dot{J}_\mathrm{torque}$):
% We can additionally consider angular momentum loss through magnetic braking ($\dot{J}_\mathrm{MB}$), which was invoked by the models of \citet{Belloni2023}.
\begin{equation}
\dot{J}_\mathrm{orbit} + \dot{J}_\mathrm{spin} = \dot{J}_\mathrm{GW} + \dot{J}_\mathrm{MB} + \dot{J}_\mathrm{eject} + \dot{J}_\mathrm{stream} + \dot{J}_\mathrm{torque}.
\label{eq:jdot}
\end{equation}

We assume that the donor is tidally locked ($\dot{J}_\mathrm{spin} = 0$) and mass transfer is conservative ($\dot{J}_\mathrm{eject} = 0$).
Following \citet{Marsh2004}, we further assume that $\dot{J}_\mathrm{stream} = - \dot{J}_\mathrm{torque}$ (in other words, the spin of the accretor is in a state of equilibrium).
\citet{Marsh2004} argued that this is a necessary condition for the survival of a binary with short orbital periods, where the radius of the accretor is a substantial fraction of the orbital separation.
Upper limits on the accretor spin derived by \citet{Kupfer2016b} are consistent with an approximately tidally locked accretor.
This being said, the arguments of \citet{Marsh2004} are mostly applicable over secular timescales.
Any short-term deviation in $\dot{J}_{\rm stream}$ may not be matched by an immediate change in $\dot{J}_{\rm torque}$, such that the accretor may serve as a short-term angular momentum sink.
This will be discussed further in the following sections.

In the situation where magnetic braking is negligible, the above simplifications give us $\dot{J}_\mathrm{orbit} \approx \dot{J}_\mathrm{GW}$.
The angular momentum of the  binary system can be expressed as 
\begin{equation}
J^3 = \frac{G^2 P}{2 \pi} \frac{(M_1 M_2)^3}{M_1 + M_2},
\end{equation}
where $M_1$ and $M_2$ are the masses of the primary and secondary stars, and $G$ is the gravitational constant.
From this comes the relation
\begin{equation}
\frac{\dot{P}}{P} = 3 \left[ \frac{\dot{J}}{J} + \frac{\dot{M}}{M_2} (1 - q) \right],
\end{equation}
where $q = M_2 / M_1$ is the mass ratio and $\dot{M} > 0$ is the mass transfer rate.
The change in angular momentum of a binary due to the emission of gravitational waves can be characterised \citep{Peters1964,Landau1971} as
\begin{equation}
\left(\frac{\dot{J}}{J}\right)_\mathrm{GW} = - \frac{32}{5} \frac{G^3}{c^5} \frac{M_1 M_2 (M_1 + M_2)}{a^4},
\end{equation}
where $a$ is the orbital separation and $c$ is the speed of light.
From these equations, Kepler's law, and the Roche lobe radius \citep{Paczynski1971}
\begin{equation}
\frac{R_L}{a} = 0.46 \left( \frac{M_2}{M_1 + M_2} \right)^3,
\end{equation}
the secular mass transfer rate can be shown to be \citep[e.g.][]{Deloye2007,Chakraborty2024}
\begin{equation}
\left(\frac{-\dot{M}}{M_2}\right)_\mathrm{sec} = \left(\frac{\dot{J}}{J}\right)_\mathrm{GW} \left(\frac{1}{5/6 + \xi/2 - q}\right),
\label{eq:mdot}
\end{equation}
and the secular orbital period evolution is
\begin{equation}
\left(\frac{\dot{P}}{P}\right)_\mathrm{sec} = 3 \left(\frac{\dot{J}}{J}\right)_\mathrm{GW} \left( \frac{\xi - 1/3}{\xi + 5/3 - 2q}\right).
\label{eq:pdot}
\end{equation}
Here we have parametrised the response of the donor radius ($R_2$) to mass loss with $\xi = d \log(R_2) / d \log(M_2)$.\footnote{
\citet{Chakraborty2024} expressed this somewhat differently, by separating the donor's short-term adiabatic response to mass loss from its longer-term evolutionary changes.
In our approach, a single parameter $\xi$, which can be calculated from evolutionary models, includes both of these donor changes.
}
% The parameter $\xi$ is a valuable descriptor of donor structure which will be discussed further in the following sections.

For binary systems with measured component masses and orbital periods, the secular \pdot\ can thus be estimated under the assumptions mentioned above and with an assumed value of $\xi$.
The typical mass-radius relation for fully-degenerate white dwarfs without accounting for mass loss would give $\xi = -1/3$.
Detailed models of the donor structure and its response to mass loss instead typically find values in the region of $\xi \approx -0.2$ for donors undergoing adiabatic expansion, which is true for most of the AM\,CVn binary period range away from the period bounce \citep[e.g.][]{Savonije1986,Deloye2007,Wong2021,Belloni2023}.
We therefore take  $\xi = -0.2$ as our default value in the following sections.
During the period bounce, or phases of donor contraction such as those described by \citet{Deloye2007}, $\xi$ may be less negative or even positive.

It is also possible to estimate $\xi$ based on an \mdot\ measurement.
\citet{Ramsay2018a} estimated the instantaneous mass transfer rate as $\log (\dot{M}/M_\odot \text{yr}^{-1}) = -10.74 \pm 0.07$ for Gaia14aae, based on the \textit{Gaia} absolute magnitude of the bright spot.
Using Equation~\ref{eq:mdot}, this would imply $\xi = -0.3 \pm 0.3$, consistent with our assumed default value of $-0.2$.
The uncertainty on the parallax of YZ\,LMi makes it impossible to place a meaningful constraint on \mdot.

In situations where magnetic braking dominates the AML from the binary, Equation~\ref{eq:jdot} instead simplifies to $\dot{J}_\mathrm{orbit} \approx \dot{J}_\mathrm{MB}$.
A wide range of prescriptions for magnetic braking are used across different fields \citep[e.g.][]{Rappaport1983,Sills2000,Van2019,El-Badry2022b}.
% Several recent works have investigated the effect of enhanced magnetic braking on the evolution of AM\,CVn binaries \citep{Belloni2023,Sarkar2023a,Sarkar2023}.
In this paper we will focus on the model tracks of \citet{Belloni2023}, who implemented the CARB prescription of magnetic braking \citep[][]{Van2019}.
Under this prescription, $\dot{J}_\mathrm{MB}$ depends on the stellar radius, spin period, mass loss rate to wind, and convective turnover timescale.
The convective turnover timescale requires the convective velocity and the size of the convective envelope, which can be calculated numerically through evolutionary models.
We refer readers to \citet{Belloni2023} for details of the calculation of $\dot{J}_{\rm MB}$ in AM\,CVn binaries.

\subsection{Known issues with period changes in other binary classes}
\label{sec:discussion:cvs}

\begin{figure}
\includegraphics[width=\columnwidth]{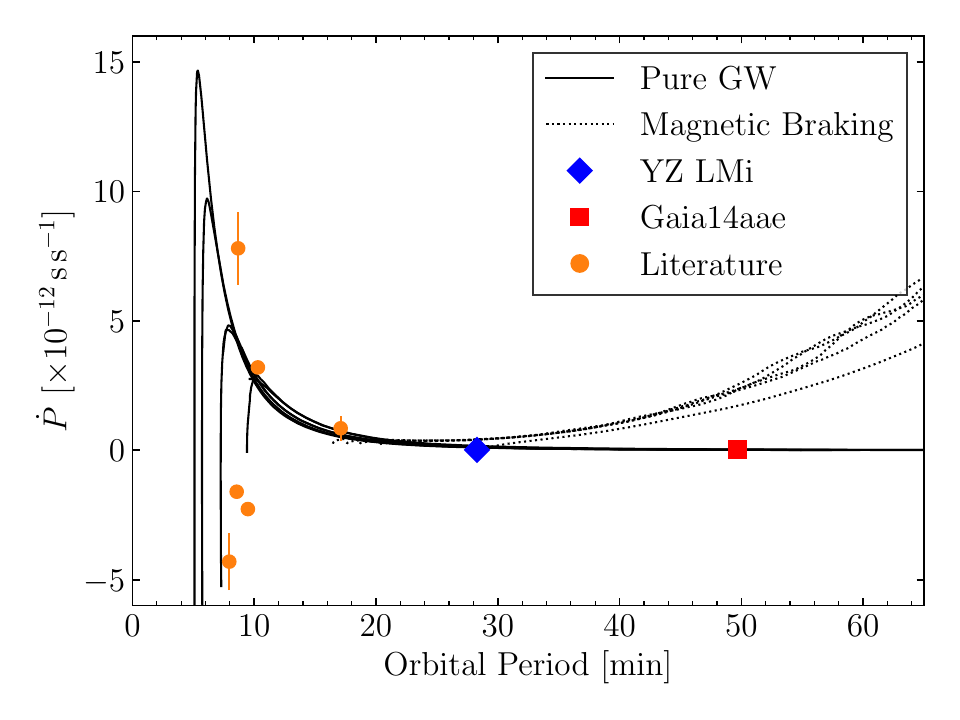}
\caption{Predicted values of \pdot\ from both sets of model tracks, compared to our measurements and those available from the literature (Table~\ref{tab:literature}).
All measured values have uncertainties, but in some cases (including YZ\,LMi and Gaia14aae) these uncertainties are smaller than the symbol size.
The very negative \pdot\ of HM\,Cnc is outside the bounds of this plot.
For model tracks from \citet{Belloni2023}, we show here only the post-bounce evolution, which is strongly influenced by magnetic braking. Their pre-bounce evolution, which is dominated by GW radiation, will be similar to the tracks of \citet{Wong2021} for similar component masses and orbital periods.
The large range of \pdot\ at orbital periods $< 20$\,min and the small \pdot\ at longer orbital periods is broadly consistent with GW emission, while the small \pdot\ of Gaia14aae is strongly inconsistent with magnetic braking models.
}
\label{fig:pdot-p}
\end{figure}

A number of short-period detached DWD binaries show a constant value of \pdot, which is generally consistent with GW emission \citep{Hermes2012,Burdge2019,Burdge2019b,Burdge2020}.
However, the picture for mass-transferring binaries is not so simple.
Among CVs, many systems do not show a constant \pdot, but instead show a periodic or quasi-periodic signature with a timescale of several decades \citep[e.g.][]{Borges2008,Qian2015,Han2015,Bours2016a,Han2017,Patterson2018,Court2019}.
\citet{Patterson2018} cautioned that, in almost all cases, the proposed periodic timescales are similar to the timescales of observation and so the veracity of the claimed period is uncertain.
\citet[][their figures 8 and 9]{Court2019} show how the suggested period of the proposed third body in Z\,Cha has regularly been forced to change with the addition of new data, perhaps a warning sign of aperiodicity.

There are two commonly discussed explanations for these (quasi-)periodic signatures: the presence of a third body, or the redistribution of angular momentum within the magnetically active donor star \citep[the Applegate mechanism:][]{Applegate1992,Watson2010}.
\citet{Patterson2018} argue the case for the Applegate mechanism in CVs, due to similarities between CV donors and the M dwarf components in post-common envelope binaries (PCEBs).
Among PCEBs, third objects were also once widely claimed \citep[e.g.][]{Qian2009,Qian2010}, but more recent observations have departed from periodic variations, leading to a general preference instead for the Applegate mechanism \citep[e.g.][]{Bours2016a,Yates2026}.
% \cmnt{add Yates2026 when published, including Yates' consistency with Applegate}
The Applegate mechanism is not without its own issues, however, as current models predict a smaller energy budget from this mechanism than would be required for the observed orbital period changes \citep[e.g.][]{Yates2026}.
In the formalism of Equation~\ref{eq:jdot}, the Applegate mechanism would drive changes in $\dot{J}_{\rm spin}$.
% It is unclear whether a mechanism such as Applegate  should affect AM\,CVn-type systems, whose donors

Even among CVs that have a measured, constant \pdot, there are issues reconciling those measurements with models.
\citet{Schaefer2024} compiled 52 measurements of \pdot\ from eclipsing CVs, and showed significant disagreements with the common model of magnetic braking in CVs. 
Their measurements scatter by up to $\dot{P} \sim \pm 10^{-9} {\rm s \, s}^{-1}$, compared to predicted values of $\sim - 10^{-11} {\rm s \, s}^{-1}$.
\citet{King2024} argue that this disagreement represents a more fundamental issue with the \pdot\ measurements in mass-transferring systems, which has been discussed since at least \citet{Pringle1975} and \citet{Ritter1988} -- see also section 4.4 of \citet{FKR}.
The issue is that a number of assumptions made in the derivation of the secular evolution of the system are potentially subject to short-term fluctuations of unknown amplitude.
These fluctuations may be driven by variations in the stellar wind, magnetic cycles, oscillations in the stellar envelope, or various other effects.
This leads to short-term variations in \mdot\ and in turn to variations in \pdot, which are assumed to average out over secular timescales but may dominate measurements on human timescales. 
In the formalism of Equation~\ref{eq:jdot}, this is equivalent to short-term changes in $\dot{J}_{\rm stream}$ which are not immediately corrected by $\dot{J}_{\rm torque}$, allowing angular momentum to be temporarily stored in or borrowed from the spin of the accretor.

In spite of these serious caveats, we will proceed for now by interpreting our measured \pdot\ as the secular value. 
We will come back to the question of short-term deviations from the secular value in Section~\ref{sec:discussion:short-term}.

\subsection{Pure GW or enhanced magnetic braking?}
\label{sec:discussion:compare-models}

\begin{figure}
\includegraphics[width=\columnwidth]{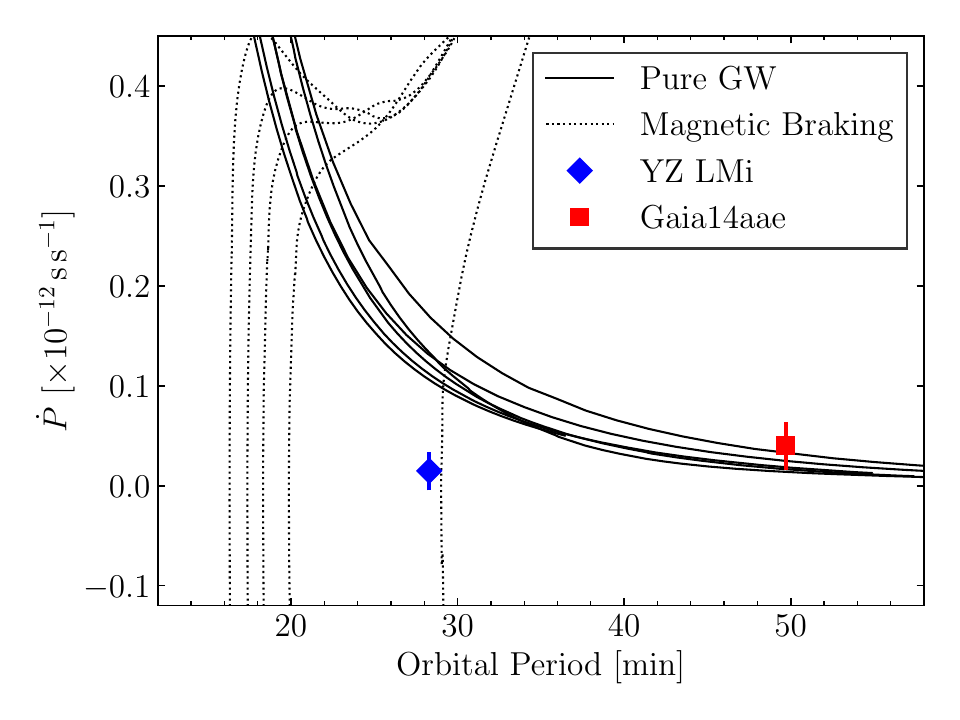}
\caption{An zoom-in of Fig.~\ref{fig:pdot-p}, showing the disagreement between the \pdot\ of YZ\,LMi and pure-GW predictions.
Here we show both pre- and post-bounce evolution for the magnetic braking tracks.
One of the tracks with magnetic braking undergoes a period bounce at approximately the orbital period of YZ\,LMi, and is therefore able to match its \pdot.
We discuss this in Section~\ref{sec:discussion:period-bounce}.
}
\label{fig:pdot-p-zoom}
\end{figure}

\begin{figure*}
\includegraphics[width=500pt]{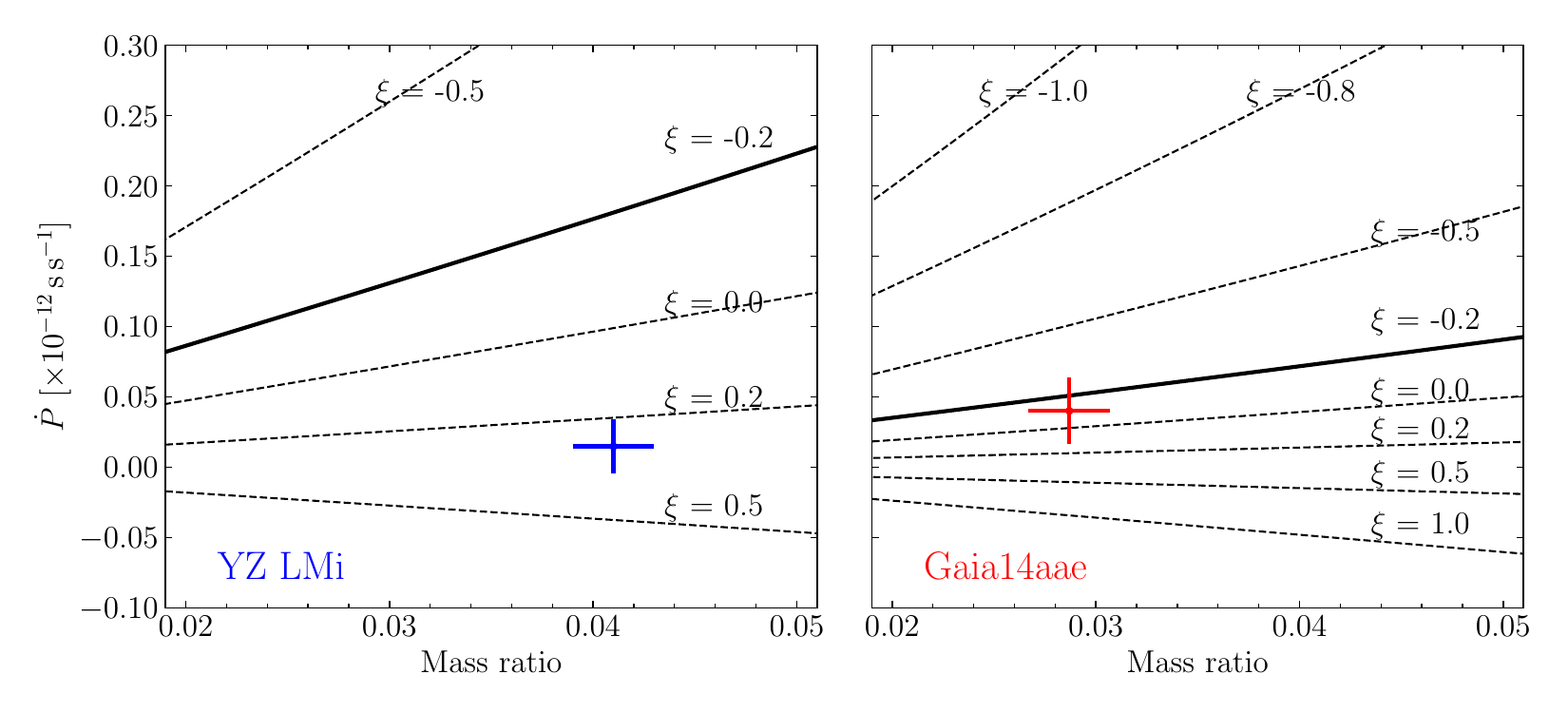}
\caption{Coloured crosses show the measured orbital mass ratios and orbital period derivatives for both systems.
We also plot predicted tracks in this phase space using Equation~\ref{eq:pdot} for a range of $\xi$ values, calculated at the orbital period of each system.
The implied $\xi$ of YZ\,LMi is positive, suggesting a donor that is contracting in response to mass loss.
The implied $\xi$ of Gaia14aae has larger uncertainties, and is consistent with either the expected value of $-0.2$ or with a positive value.
}
\label{fig:pdot-xi}
\end{figure*}

\begin{figure}
\includegraphics[width=\columnwidth]{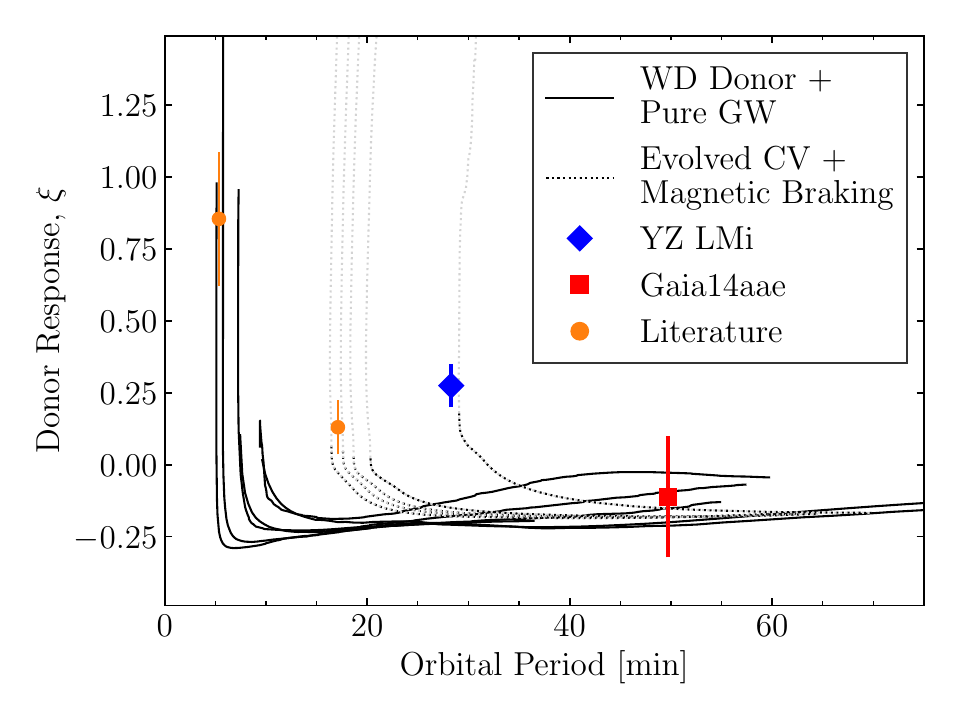}
\caption{Donor response to mass loss ($\xi$) as a function of orbital period, as predicted by both sets of evolutionary tracks, compared to the $\xi$ values for our targets and those from Table~\ref{tab:literature} implied by the measured \pdot\ values.
Tracks from \citet{Belloni2023} are plotted in a fainter shade of grey where the hydrogen fraction in the accreted material is $> 10^{-3}$.
We note that the estimated $\xi$ for HM\,Cnc (the leftmost data point) is that implied by the mass ratio of \citet{Roelofs2010}, whereas the mass ratio of \citet{Munday2023} would give a substantially larger $\xi$ beyond the field of this plot.
Pure GW model tracks are not able to replicate the positive \pdot\ of YZ\,LMi at its orbital period, while models with magnetic braking only reproduce these \pdot\ values if the target is undergoing a period bounce.
}
\label{fig:xi-p}
\end{figure}

Under the assumption that AML is dominated by GW radiation, Equation~\ref{eq:pdot} predicts $\dot{P} = (1.8 \pm 0.2) \times10^{-13} {\rm s \, s}^{-1}$ %5.1 \mu \mathrm{s \, yr}^{-1}$ 
for YZ\,LMi and 
$(5.2 \pm 0.6) \times10^{-14} {\rm s \, s}^{-1}$
% $1.7 \mu \mathrm{s \, yr}^{-1}$ 
for Gaia14aae, where we have taken physical parameters from \citet{Copperwheat2011} and \citet{Green2018a} and assumed $\xi = -0.2$.
The expected departures from a linear ephemeris due to these \pdot\ values are shown in Fig.~\ref{fig:ephemeris}.
For Gaia14aae the prediction is within $2\sigma$ of the measured value, while for YZ\,LMi it differs by more than $5\sigma$.

It is also possible to predict \pdot\ numerically from evolutionary models.
From MESA model tracks calculated by \citet[][]{Wong2021} and \citet{Belloni2023}, we calculate \pdot\ by numerically differentiating \porb\ with respect to time.
The resulting relationships between \pdot\ and \porb\ for both sets of tracks are shown in Fig.~\ref{fig:pdot-p}, compared to our measurements and those from the literature shown in Table~\ref{tab:literature}. 
% The values at larger \porb\ are the best discriminators between the competing models of AML.

At $P \lesssim 25$\,min, both sets of model tracks are dominated by GW radiation for post-bounce systems.\footnote{The model tracks shown here from \citet{Belloni2023} do not reach $P < 20$\,min, but other tracks that undergo period bounces at shorter periods agree approximately with the tracks of \citet{Wong2021} at those periods.}
The strongest difference between pure-GW models of \citet{Wong2021} and the CARB models of \citet{Belloni2023} occurs at orbital periods $\gtrsim 30$\,min, where CARB models predict that magnetic braking becomes the dominant source of AML.
This is a result of the surface convection zone of the donor atmosphere extending deeper within the donor, leading to a longer convective turnover time -- a prediction of the MESA models.
As discussed by \citet[][their figure 7]{Belloni2023}, this post-bounce magnetic braking is necessary to explain the bloated donor radius observed for Gaia14aae.

The overall distribution of \pdot\ measurements shows large amplitudes (both positive and negative) at short periods, and smaller amplitudes at longer periods.
The \pdot\ measurement for Gaia14aae is the best discriminator between the two AML scenarios, and strongly favours the pure GW models.
The disagreement with CARB models is difficult to reconcile unless the MESA atmospheric models of the donor are dramatically incorrect, or by dropping our interpretation that this is the secular \pdot\ (as discussed in Section~\ref{sec:discussion:short-term}).

For YZ\,LMi, interpretation is less straightforward.
In this period range, CARB magnetic braking does not substantially change the predicted \pdot, and so both model scenarios make similar predictions.
As Fig.~\ref{fig:pdot-p-zoom} shows, there is a small but significant ($5\sigma$) discrepancy between our measured \pdot\ for YZ\,LMi and the predictions of even pure GW model tracks, equivalent to the disagreement with the prediction of Equation~\ref{eq:pdot} mentioned previously.
There is one model track from \citet{Belloni2023} which does match the \pdot\ of YZ\,LMi.
That modelled system underwent a period bounce at the orbital period of YZ\,LMi, and raises the intriguing possibility that YZ\,LMi is a period-bounce system.
We discuss this in the next section.

\subsection{Is YZ\,LMi a period-bounce system?}
\label{sec:discussion:period-bounce}

\begin{figure}
\includegraphics[width=\columnwidth]{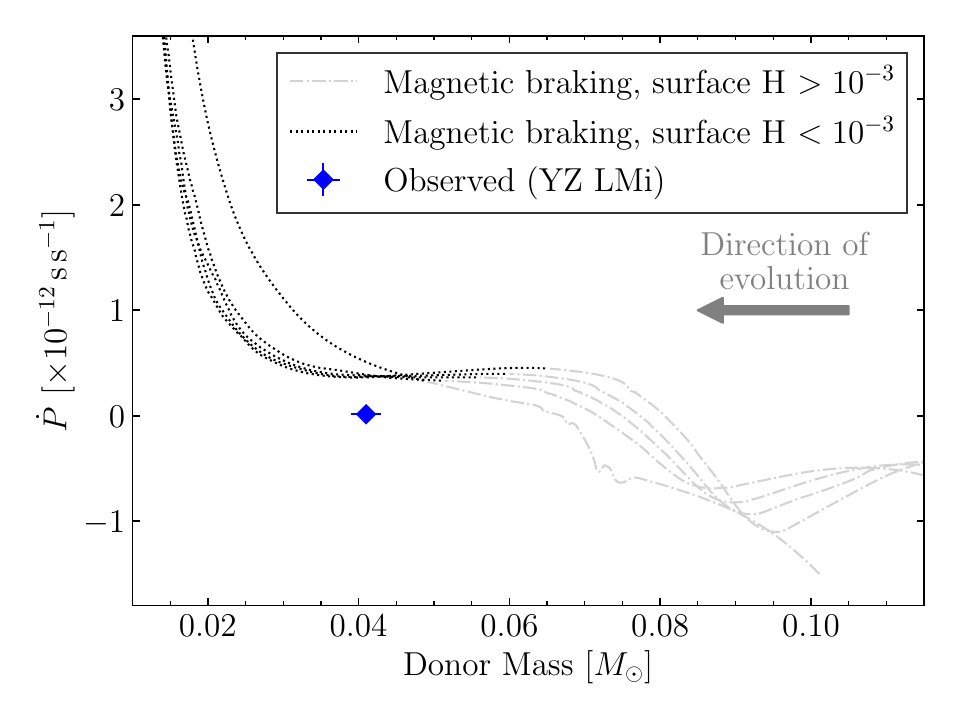}
\caption{Predicted \pdot\ and donor mass values from the evolved CV tracks of \citet{Belloni2023}, compared to the measured values for YZ\,LMi.
Although some tracks are separately able to reproduce the observed donor mass (Fig.~\ref{fig:m2-p}) and \pdot\ (Fig.~\ref{fig:pdot-p-zoom}), no track is able to reproduce both measurements at the same point in the evolution of the binary.
We also note that all model tracks continue to show detectable levels of surface hydrogen until their \pdot\ is substantially more positive than the YZ\,LMi measurement.
}
\label{fig:pbounce-m2}
\end{figure}

\begin{figure}
\includegraphics[width=\columnwidth]{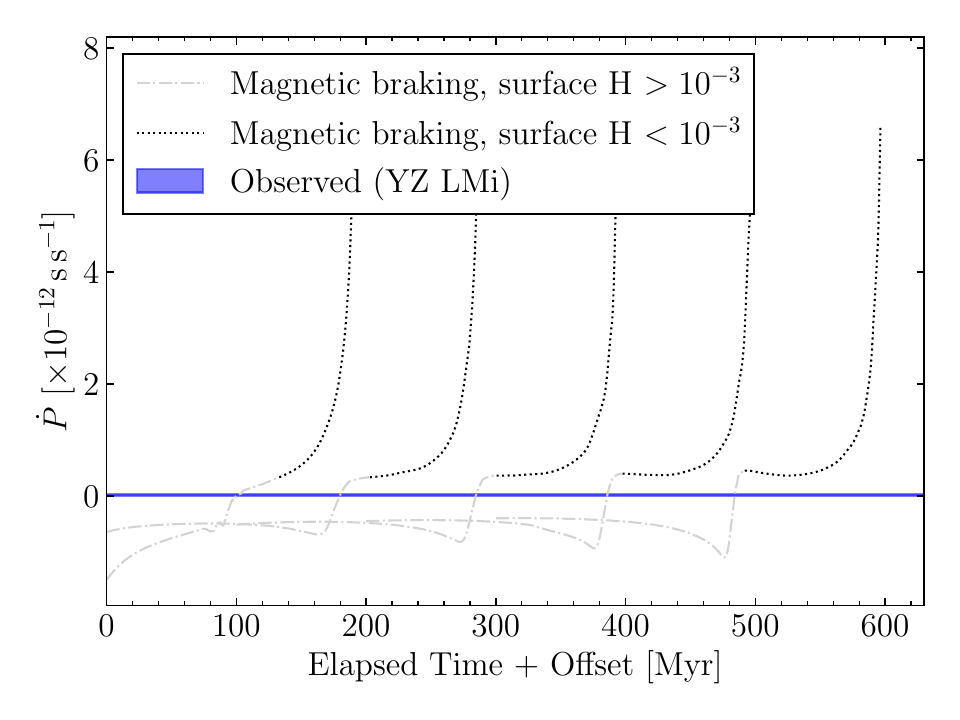}
\caption{The evolution in \pdot\ as a function of time for model tracks from \citet{Belloni2023}, compared to the $1 \sigma$ range measured for YZ\,LMi.
While all model tracks do pass through $\dot{P} = 0$, they spend only a short fraction of their observable lifespans with \pdot\ values consistent with our observations.
}
\label{fig:pbounce-time}
\end{figure}

% Under the pure GW model tracks in Fig.~\ref{fig:pdot-p}, there is a small but significant ($?\sigma$) discrepancy between our measured \porb\ for YZ\,LMi and the predictions of all model tracks.

In this section, we shall continue to interpret our measured \pdot\ values under the assumption that this is the secular \pdot.
As the measured \pdot\ for YZ\,LMi is smaller than the model prediction, it cannot be reconciled by invoking extra sources of AML. 
It is also impossible to reduce the GW-based AML without substantially changing the masses measured by \citet{Copperwheat2011}.
Therefore, the only possible reconciliation between our measured value and Equation~\ref{eq:pdot} is to change the parameter of $\xi$.

In Fig.~\ref{fig:pdot-xi}, we show the values of $\xi$ implied by our measurements: $\xi = 0.27 \pm 0.08$ for YZ\,LMi and $-0.1 \pm 0.2$ for Gaia14aae.
The positive value of $\xi$ for YZ\,LMi is not consistent with a white dwarf donor undergoing adiabatic expansion, and suggests instead a donor which is shrinking in response to mass loss.
Fig.~\ref{fig:xi-p} shows the values of $\xi$ predicted by \citet{Wong2021} and \citet{Belloni2023} for the entire period range. 
Such a positive value is not expected in this period range under the models of \citet{Wong2021}, but is predicted during a period bounce by \citet{Belloni2023}.

The natural suggestion that arises is that YZ\,LMi is a period-bounce system descended from the evolved CV track. %, illustrated by one of our model tracks from \citet[][but note that this track was selected due to its agreement with Gaia14aae, not YZ\,LMi, in period-donor mass space]{Belloni2023}.
However, we stress that there is no evolutionary track from \citet{Belloni2023} that matches both the observed \pdot\ and other observed parameters.
The model track which passes through its period bounce at the orbital period of YZ\,LMi does so with a substantially larger donor mass ($M_2 \approx 0.07 M_\odot$) than the value of $M_2 = 0.035 \pm 0.003$ measured by \citet{Copperwheat2011}.
This issue is illustrated in Fig.~\ref{fig:pbounce-m2}.
Additionally, that model track, and indeed all tracks of \citet[][see their figure 4]{Belloni2023}, maintain detectable levels of hydrogen in the accreted material until after the period bounce, compared to the non-detection of hydrogen in YZ\,LMi.
Hydrogen in AM\,CVn binaries should be detectable at abundances $\lesssim 10^{-4}$ \citep[e.g.][]{Nagel2009}, so fractions $> 10^{-3}$ can be easily ruled out with existing observations.

A second issue with the period bounce interpretation concerns timescales.
During the period bounce, the binary is evolving quickly, with a large second derivative $\Ddot{P}$.
The timescale of the period bounce is therefore rather short compared to the lifetime in which the binary is observable as an AM\,CVn system.
In Fig.~\ref{fig:pbounce-time}, we show \pdot\ as a function of elapsed time for the portion of the evolved CV tracks with $P < 70$\,min.
Each track is only consistent with our measured \pdot\ for YZ\,LMi for 0.2 to 3\,Myr out of a total lifetime of 100\,Myr.
We note that this is not a result of the CARB model implemented by \citet{Belloni2023}, as GW emission dominates the AML of each of these tracks during its period bounce.
It is therefore unlikely (though not impossible) that one of our two long-period measurements to date would happen to be caught during this stage of its evolution.

\subsection{Non-secular period changes}
\label{sec:discussion:short-term}
%
%
%\begin{figure}
%\includegraphics[width=\columnwidth]{figures/depths/output.pdf}
%\caption{The flux from various components of YZ\,LMi, as measured from the eclipse depths of each eclipse component from $g'$ observations in the 2006--2015 timespan.
%The 2012 observations, which occured several weeks after a dwarf nova outburst, are marked, and show a somewhat brighter accretion disc.
%Other components, most notably the accretor, do not show any variability over the 9 year span.
%Later observations with HiPERCAM were not included, as the $g_s$ filter showed a noticeable offset from the $g'$ measurements shown here.
%}
%\label{fig:depths}
%\end{figure}

We now return to the issues raised in Section~\ref{sec:discussion:cvs}, and challenge our assumption that the measured \pdot\ represents the secular value.
It is unclear whether the quasi-periodic eclipse variations seen in CVs should affect AM\,CVn binaries.
If the underlying mechanism is Applegate, we might assume that the donors of AM\,CVn binaries, having a very different nature, are either not affected or at least are affected to a lesser extent.
Our 20 year observational baseline for YZ\,LMi would be enough to observe changes on the typical 20--30 year timescale of CV eclipse timing variations.

The broader issue of donor atmospheric structure raised by \citet{King2024} is more challenging.
\citet{King2024} argue that the secular value of \pdot\ can only be measured on timescales sufficiently long that several scale heights of the donor's atmosphere are stripped.
The necessary timescale would be given by 
\begin{equation}
t_H = \frac{H}{| \dot{R_2} |} = \frac{H}{R_{2}} \frac{R_{2}}{ | \dot{R}_{2} | }  \approx \frac{H}{R_{2}} \frac{M_{2}}{ \dot{M} } ,
\label{eq:h-timescale}
\end{equation}
where $H$ is the atmospheric scale height of the donor and $R_2$ is the donor radius.

We estimate this timescale across the expected parameter distribution of AM\,CVn binaries.
The atmospheric scale height of the donor can be estimated as 
\begin{equation}
    H \approx \frac{ k_{\rm B} T_{\rm eff,2} }{ \mu m_{\rm p} g_2 } ,
\end{equation}
where $k_{\rm B}$ is Boltzmann's constant, $T_{\rm eff,2}$ is the effective temperature of the donor, $\mu \approx 4$ is the mean molecular weight, $m_{\rm p}$ is the proton mass, and $g_2 = G M_{2} / R_{2}^{2}$ is the surface gravity of the donor. 
Although donors in AM\,CVn binary systems at short orbital periods are hotter, they also have a higher surface gravity, leading to $H/R_{2}$ varying by order unity across the AM\,CVn population.
Taking as example values $M_{2} \approx 0.03 \, M_{\odot}$, $R_{2} \approx 0.05 \, R_{\odot}$, and $T_{\rm eff,2} \approx 2000$~K, we estimate $H/R_{2} \approx 4 \times 10^{-5}$. 

We take $\dot{M} \sim 10^{-10}\,M_{\odot}\,\rm{yr}^{-1}$, which is appropriate at $P\approx 30$~min for mass transfer driven by gravitational waves \citep[e.g.][]{Wong2021} and comparable to the instantaneous measured value for Gaia14aae of $\log (\dot{M}/M_\odot {\rm yr}^{-1}) = -10.74 \pm 0.07$ by \citet{Ramsay2018a}.
We find
\begin{equation}
    t_{\rm H} 
    \approx 10^{4}~\rm{yr} \left( \frac{H/R_{2}}{4 \times 10^{-5}} \right) \left( \frac{M_{2}}{0.03\,M_{\odot}} \right) \left( \frac{\dot{M}}{10^{-10}\,M_{\odot}\,\rm{yr}} \right)^{-1} .
\end{equation}
Our observational baseline has therefore averaged over substantially less than a scale height of the donor atmosphere for both targets.\footnote{
This timescale is only comparable to observational baselines among the very shortest-period AM\,CVn binaries.
% This timescale scales sharply with $P$, since $\dot{M} \propto P^{-4.5 - (-6)}$ for mass transfer driven by GW radiation \citep[e.g.,][]{Levitan2015,Wong2021}. The upshot is that it is easier to detect secular period changes in short-period AM~CVn binaries, especially the ones at $P \lesssim 10$~min. 
These systems have more massive donors, $M_{2} \approx 0.1\,M_{\odot}$, and much higher secular $\dot{M} \gtrsim 10^{-8}\,M_{\odot}\,\rm{yr}^{-1}$, leading to $t_{\rm H} \sim 10$--$100$\,yr. 
This, combined with the larger secular $\dot{P}$ expected at short \porb\ under pure GW radiation, may help to explain why the measured $\dot{P}$ of short-period AM\,CVn systems can be reconciled with GW expectations without invoking the deviations from secular values discussed here \citep[e.g.][]{Burdge2023,Chakraborty2024,Chickles2026}. 
}

% Taking $H/R \sim 10^{-5}$ and the mass transfer timescale $\dot{M_2} / M_2 \sim 10^7$\,yr from \citet{Wong2021}, Equation~\ref{eq:h-timescale} gives $t_H \sim 100$\,yr.
% Our observational baseline has therefore averaged over substantially less than a scale height of the donor atmosphere for both targets.

This does not mean that the \pdot\ measurements presented here are meaningless.
If short-term deviations from the secular \pdot\ can be assumed to have some characteristic amplitude, and if that amplitude is consistent between AM\,CVn binaries (as seems likely, given that their donor stars have a similar nature across the population), we would expect a sample of \pdot\ measurements across the population to follow a general trend driven by the secular evolution, with a superimposed scatter due to the short-term deviations.
If the scatter is smaller than the trend, then the trend can still be determined from a large sample of \pdot\ measurements.
Even with only two measurements, we can already make some qualitative arguments that the scatter is small.

Firstly, we consider that our two measurements are consistent with each other (within $1\sigma$) and with zero (within $2\sigma$).
From this alone, we can argue that any scatter is unlikely to be significantly larger than the measurement uncertainties.

Secondly, we note that \citet[][their figure 1]{Schaefer2024} show \pdot\ values with magnitudes much larger than ours.
Their measurements are in the range $10^{-12} \lesssim |\dot{P}| \lesssim 10^{-9} \, {\rm s \, s}^{-1}$, compared to model predictions $|\dot{P}| < 10^{-11}  \, {\rm s \, s}^{-1}$.
Only two of their 52 measurements have an amplitude $|\dot{P}| < 10^{-12} \, {\rm s \, s}^{-1}$, and none have $|\dot{P}| < 10^{-13} \, {\rm s \, s}^{-1}$.
By contrast, all AM\,CVn measurements except for the shortest-period system, HM\,Cnc, have $|\dot{P}| < 10^{-12} \, {\rm s \, s}^{-1}$, and both long-period measurements presented in this work have amplitudes $|\dot{P}| < 10^{-13} \, {\rm s \, s}^{-1}$.
Whatever the nature of the scatter among CV measurements is, it seems clear that AM\,CVn binaries do not experience scatter of a similar magnitude.
This is not necessarily surprising: the \mbox{(semi-)degenerate} and helium-dominated donors of AM\,CVn systems are quite different from the M-dwarf donors of hydrogen-rich CVs, and so there is no reason to expect similar physical mechanisms to be present.

Thirdly, we return to the pure-GW and CARB models of AM\,CVn evolution discussed in Section~\ref{sec:discussion:compare-models}, and calculate the amplitude of short-term deviation, $\Delta \dot{P}$, from the secular \pdot\ required to produce our measured \pdot\ values under each scenario.
In the case of pure GW radiation, reconciling our \pdot\ value for YZ\,LMi with the prediction of Equation~\ref{eq:pdot} would require a short-term deviation of $\Delta\dot{P} = (-1.3 \pm 0.3)\times10^{-13} \, {\rm s \, s}^{-1}$. 
Maintaining the agreement of Gaia14aae would require $\Delta\dot{P} = (-0.3 \pm 0.3) \times10^{-13} \, {\rm s \, s}^{-1}$.
In other words, a scenario of pure GW combined with relatively small-amplitude deviations ($\lesssim 10^{-13} {\rm s \, s}^{-1}$) readily explains both measurements.

In the CARB scenario, YZ\,LMi would require a similar $\Delta\dot{P}$ to the pure GW scenario, while reconciling Gaia14aae would require much larger short-term deviations of, depending on the chosen model track, $\Delta\dot{P} = -1.7 \times10^{-12} \, {\rm s \, s}^{-1}$ to $-2.6 \times10^{-12} \, {\rm s \, s}^{-1}$.
% Therefore, reconciling the CARB scenario with our measurements would require deviations of amplitude $\sim 10^{-12} \, {\rm s \, s}^{-1}$ to be possible.
Two substantial coincidences are then necessary to explain our non-detections of significant \pdot\ under this scenario.
Firstly, because the amplitude of the short-term deviation in Gaia14aae must be an order of magnitude larger than the uncertainties on our measurement, its amplitude must be fine-tuned to a level of a few per cent in order to make the observable \pdot\ consistent with zero.
Secondly, the amplitude of short-term deviation in Gaia14aae must be an order of magnitude larger than that of YZ\,LMi, in order to make both measurements consistent with zero despite very different secular values.
% While we cannot disprove this scenario, the principle of Ockham's razor favours the pure GW scenario given the currently available data.
%
%
%Finally, we note that short-term deviations from the secular \pdot\ should be matched by changes in \mdot.
%%From Equations~\ref{eq:mdot} and \ref{eq:pdot}, we can estimate that the 
%\citet{Levitan2015} previously observed an apparent change in 

While the present data are not sufficient to disprove alternative scenarios, the scenario that explains the \pdot\ measurements with the least contrivence is one of pure GW combined with relatively small amplitude ($\Delta \dot{P} \lesssim 10^{-13} \, {\rm s \, s}^{-1}$) deviations from the secular \pdot\ over timescales significantly longer than two decades.
This discussion is naturally limited by the small number of \pdot\ measurements at long periods.
Further measurements of \pdot\ will enable a more thorough characterisation of the scatter induced by short-term variations, and hence of the underlying \pdot\ trend.

\cmnt{I was planning to say something here about changes in Mdot and outburst rate, but I think it's a complicated argument to make and doesn't add much to the overall picture.
The summary is -- Short-term changes in Pdot should lead to changes in Mdot. 
No strong evidence that outburst rate changes. Component magnitudes don't change. Therefore I don't think Mdot has changed over the timescale of our observations.
But, as the O-C plot is a flat line, I think we can already rule out changes in Pdot over timescales $<10$ years anyway, and I think that argument is stronger than trying to guess at Mdot.
}

\section{Conclusions}

In this work we have presented a search for orbital period evolution in two long-period AM\,CVn binary systems, YZ\,LMi and Gaia14aae.
Despite 447 and 181 mid-eclipse time measurements across observational baselines of 19 and 10 years, respectively, we do not detect any significant orbital period evolution.
Our $3\sigma$ upper limits on \pdot\ are $1.1\times10^{-13} \, {\rm s \, s}^{-1}$ and $9.8\times10^{-14} \, {\rm s \, s}^{-1}$ respectively.

These upper limits place constraints on the sources of AML in evolutionary models.
We compare our upper limits to two scenarios, where AML is dominated by GW emission \citep{Wong2021} or magnetic braking under the CARB prescription \citep{Belloni2023}.
Our upper limit on \pdot\ for Gaia14aae is consistent with GW predictions but an order of magnitude smaller than predictions under the CARB model.
Our upper limit on \pdot\ for YZ\,LMi shows a small ($\Delta\dot{P} \approx 10^{-13} \, {\rm s \, s}^{-1}$) but significant ($5\sigma$) disagreement with predictions under both scenarios.

We consider the large-amplitude, short-term deviations from secular \pdot\ seen among hydrogen-dominated CVs, \review{which often show quasi-periodic variations in orbital period over timescales of decades \citep[e.g.][]{Patterson2018,Court2019} and where most measured values of \pdot\ disagree substantially with predicted values \citep[e.g.][]{Schaefer2024,King2024}. 
These issues highlight the importance of caution when interpreting \pdot\ measurements in mass-transferring binary systems.
Nevertheless,} we present empirically-driven arguments that the short-term deviations among AM\,CVn binaries, if present, appear to be substantially smaller in amplitude. 
% We argue that deviations of amplitude $|\Delta \dot{P}| \lesssim 10^{-13}$ are most likely, while amplitudes $\lesssim 10^{-12}$ are also possible.
Deviations of amplitude $|\Delta \dot{P}| \gtrsim 10^{-12} \, {\rm s \, s}^{-1}$ do not appear to be present in any of the AM\,CVn binaries observed so far, in contrast to the deviations of $10^{-12} \lesssim |\Delta \dot{P}|\lesssim 10^{-9} \, {\rm s \, s}^{-1}$ that are observed in hydrogen CVs.

The simplest explanation for our results is a scenario in which the secular AML from the binary is dominated by (or at least not substantially larger than) GW radiation, while short-term deviations with amplitudes $\lesssim 10^{-13} \, {\rm s \, s}^{-1}$ induce a scatter around the secular \pdot\ when observed on human timescales.
An interpretation in which the much larger secular \pdot\ predicted for Gaia14aae under the CARB model is hidden by a short-term deviation is possible, but requires that deviation to be fine-tuned to the level of a few per cent in order to reproduce our non-detection.
We have also considered a scenario in which the \pdot\ measured for YZ\,LMi is the secular \pdot\ and YZ\,LMi is currently undergoing a period bounce, but we have shown that its combination of \pdot, donor mass, and composition are not reproduced by any model track.

Further study of long-period AM\,CVn binaries will allow a more thorough 
\review{investigation of orbital period changes in this binary class.}
At least ten long-period, eclipsing AM\,CVn binaries have been discovered in recent years \citep[e.g.][]{vanRoestel2022,Rodriguez2023,Khalil2024}, and measurements of \pdot\ of comparable precision to ours will become possible for these systems in the near future.
\review{Such measurements, if performed with consideration of potential systematic issues such as those present in hydrogen CVs, have the potential both to further constrain the strength of AML in AM\,CVn binaries and to address some of the uncertainties considered in this work.
In particular, it will be valuable to understand whether the quasi-periodic changes in orbital period seen in hydrogen CVs are also present in AM\,CVn binaries, and whether other AM\,CVn binaries show deviations from the secular \pdot\ of a similar amplitude to that which we have invoked to explain YZ\,LMi.}
We hope that the considerations laid out in this work can inform the interpretation of those measurements.

% characterisation of both the underlying signal (the trend in secular \pdot\ across the population) and the scatter on that signal (the short-term deviations in \pdot\ that affect individual measurements).
% At least ten long-period, eclipsing AM\,CVn binaries have been discovered in recent years \citep[e.g.][]{vanRoestel2022,Rodriguez2023,Khalil2024}.
% Measurements of \pdot\ of comparable precision to ours will become possible for these systems in the near future.
% We hope that the considerations laid out in this work can inform the interpretation of those measurements.

\section*{Acknowledgements}

We are grateful to the anonymous referee for their insightful comments. 
MJG thanks Mitch Begelman and the JILA department at the University of Colorado, Boulder, for providing office space at which much of this paper was written.

This work is supported in part by the NASA under grants 80NSSC24K0436, 80NSSC22K0479, and 80NSSC24K0380, and the NSF under grant  AST-2508429.
VSD and HiPERCAM are funded by the Science and Technology Facilities Council (grant ST/Z000033/1).
IP acknowledges support from the Royal Society through a University Research Fellowship (URF\textbackslash R1\textbackslash 231496). This project has received funding from the European Research Council under the European Union’s Horizon 2020 research and innovation programme (grant agreement numbers 101002408 – MOS100PC). CMC receives funding from UKRI grant numbers ST/X005933/1 and ST/W001934/1.

\cmnt{Feel free to add your own acknowledgements}

This article is based in part on observations made in the Observatorios de Canarias del IAC with the the William Herschel Telescope (WHT) operated on the island of La Palma by the Isaac Newton Group (ING) in the Observatorio del Roque de los Muchachos.
It is also based in part on observations made with the Gran Telescopio Canarias (GTC) under proposal ID GTC18-24A, installed at the Spanish Observatorio del Roque de los Muchachos of the Instituto de Astrofísica de Canarias, in the island of La Palma.
Further data were obtained using the 2.4 m Thai National Telescope (TNT) operated by the National Astronomy Research Institute of Thailand (NARIT), and the 200-inch Hale Telescope at Palomar Observatory operated by the California Institute of Technology.

Software packages used in this work include the \software{ultracam} and \software{hipercam} reduction pipelines, \software{lcurve} \citep{Copperwheat2010}, \software{numpy}, \software{astropy}, \software{matplotlib}, and \software{emcee} \citep{Foreman-Mackey2013}.

\section*{Data Availability}

Raw and reduced HiPERCAM data are available through the GTC public archive.
All other data analysed in this work can be made available upon reasonable request to the authors.

%%%%%%%%%%%%%%%%%%%%%%%%%%%%%%%%%%%%%%%%%%%%%%%%%%



\bibliographystyle{mnras}
\bibliography{refs} % if your bibtex file is called example.bib

%\begin{thebibliography}{99}
%\end{thebibliography}

%%%%%%%%%%%%%%%%%%%%%%%%%%%%%%%%%%%%%%%%%%%%%%%%%%

%%%%%%%%%%%%%%%%% APPENDICES %%%%%%%%%%%%%%%%%%%%%

\appendix

\section{Summary of Observations}

In Tables~\ref{tab:observations-yzlmi-1}--\ref{tab:observations-gaia14aae}, we give the dates and observational set-ups for all data collected for this work.

\begin{table}
\caption{Observations of YZ\,LMi with ULTRACAM and HiPERCAM. 
``Exp.'' represents the exposure time.
For ULTRACAM observations, the \filu\ exposure time was $3\times$ that of the other bands.
For HiPERCAM observations, different exposure times were used in different filters, and these are listed separately although the observations were simultaneous.
Parentheses around the observation date indicates that the observations coincided with a dwarf nova outburst.
}
\begin{tabular}{lcccc}
\label{tab:observations-yzlmi-1}
Telescope & Date & Filter & Exp.     & Num. \\
+ instrument           &      &        & [s]          & eclipses \\
\hline
% \textit{YZ LMi}\\
WHT +      &    2006 Mar 01    &      \filu\filg\filr      &                3.0   &     14 \\
ULTRACAM                   &    2006 Mar 02    &      \filu\filg\filr      &                3.0   &     20 \\
                   &    2006 Mar 03    &      \filu\filg\filr      &                3.0   &     17 \\
                   &    2009 Jan 01    &      \filu\filg\filr                  &    1.8   &     3 \\
                   &    2009 Jan 02    &      \filu\filg\filr      &                1.8   &     18 \\
                   &    2009 Jan 03    &      \filu\filg\filr      &                1.8   &     11 \\
                   &    (2012 Jan 14)    &      \filu\filg\filr                  &    6.0   &     5 \\
                   &    (2012 Jan 17)    &      \filu\filg\filr                  &    4.0   &     3 \\
                   &    (2012 Jan 18)    &      \filu\filg\filr      &                4.5   &     11 \\
                   &    (2012 Jan 19)    &      \filu\filg\filr                  &    4.5   &     2 \\
                   &    (2012 Jan 20)    &      \filu\filg\filr                  &    4.5   &     3 \\
                   &    (2012 Jan 21)    &      \filu\filg\filr      &                2.0   &     10 \\
                   &    (2012 Jan 22)    &      \filu\filg\filr      &                2.0   &     14 \\
                   &    (2012 Jan 28)    &      \filu\filg\filr                  &    2.0   &     4\\
                   &    2012 Feb 01    &      \filu\filg\filr                  &    2.0   &     6\\
                   &    2012 Feb 03    &      \filu\filg\filr                  &    2.0   &     8\\
                   &    2013 Dec 30    &      \filu\filg\filr                  &    4.0   &     7\\
                   &    2013 Dec 31    &      \filu\filg\filr                  &    4.0   &     11\\
                   &    2015 Jan 14    &      \filu\filg\fili                  &    4.0   &     4\\
                   &    2015 Jan 15    &      \filu\filg\filr                  &    4.0   &     6\\
\\
GTC +      &    2024 Mar 06    &    \filgs   &  2.5 & 2  \\
HiPERCAM      &       &     \filrs\filis      &  5.0 & 2  \\
    &        &      \filus\filzs   &  7.5 & 2  \\
\hline
\end{tabular}
\end{table}

\begin{table}
\caption{Observations of YZ\,LMi with ULTRASPEC.
``Exp.'' represents the exposure time. On some occasions, multiple exposure times were used between different runs on a single night.
Parentheses around the observation date indicates that the observations coincided with a dwarf nova outburst.
}
\begin{tabular}{lcccc}
\label{tab:observations-yzlmi-2}
Telescope & Date & Filter & Exp.     & Num. \\
+ instrument           &      &        & [s]          & eclipses \\
\hline
TNT +     &    2017 Feb 18    &      \filkg                           &    5.0   &     2\\
ULTRASPEC                   &    2017 Feb 20    &      \filkg                           &    5.0   &     2\\
                   &    2017 Feb 22    &      \filkg                           &    5.0   &     3\\
                   &    2017 Feb 24    &      \filkg                           &    5.0   &     2\\
                   &    2017 Dec 05    &      \filkg                           &    5.0   &     2\\
                   &    (2017 Dec 12)    &      \filkg                           &    6.3   &     1\\
                   &    2017 Dec 14    &      \filkg                           &    8.0   &     2\\
                   &    2017 Dec 15    &      \filkg                           &    8.0   &     1\\
                   &    2018 Feb 07    &      \filkg                           &    5.0   &     1\\
                   &    2018 Feb 09    &      \filkg                           &    5.5   &     3\\
                   &    2018 Feb 10    &      \filkg                           &    5.0   &     1\\
                   &    2019 Dec 08    &      \filkg                           &    5.0   &     5\\
                   &    2020 Jan 17    &      \filkg                           &    8.0   &     9\\
                   &    2020 Jan 18    &      \filkg                           &    10.0  &     3\\
                   &    2020 Jan 19    &      \filkg                           &    10.0  &     3\\
                   &    2020 Jan 25    &      \filkg                           &    6.0   &     2\\
                   &    2020 Feb 01    &      \filkg                           &    6.0   &     2\\
                   &    2020 Feb 03    &      \filkg                           &    6.0   &     2\\           
                   &	2020 Dec 21		&	  \filkg						   &    15.0    &     2\\      
                   &	2021 Jan 09    &	  \filkg						   &    8.6    &     2\\      
                   &	2021 Feb 01		&	  \filkg						   &    6.6    &     1\\      
                   &	2021 Feb 15		&	  \filkg						   &    6.6    &     1\\      
                   &	2021 Mar 09		&	  \filkg						   &    6.3    &     1\\
& 2023 Jan 13 & \filkg & 7.8, 9.8 & 2 \\
& 2023 Jan 14 & \filkg & 7.3, 8.8 & 2\\
& 2023 Feb 01 & \filkg & 6.0 & 1\\
& 2023 Feb 08 & \filkg & 20.0, 14.0 & 3\\
& 2023 Feb 09 & \filkg & 14.0, 20.0 & 3\\
& 2023 Feb 10 & \filkg & 10.0 & 2\\
& 2023 Feb 22 & \filkg & 8.6 & 2\\
& (2023 Mar 14) & \filkg & 10.0 & 1\\
& (2023 Mar 15) & \filkg & 10.0 & 2\\
& 2023 Mar 16 & \filkg & 10.0 & 1\\
& 2023 Nov 18 & \filkg & 10.0 & 1\\
& 2023 Nov 19 & \filkg & 10.0 & 1\\
& 2023 Nov 20 & \filkg & 10.0 & 1\\
& 2024 Feb 06 & \filkg & 6.5 & 1\\
& 2024 Feb 07 & \filkg & 6.0 & 3\\
& 2024 Feb 14 & \filkg & 6.0 & 3\\
& 2024 Feb 28 & \filkg & 3.0 & 1\\
& 2024 Mar 11 & \filkg & 6.0 & 1\\
\hline
\end{tabular}
\end{table}

\begin{table}
\caption{Observations of Gaia14aae with all instruments.
``Exp.'' represents the exposure time.
For ULTRACAM observations, the \filu\ exposure time was $3\times$ that of the other bands.
For HiPERCAM observations, different exposure times were used in different filters, and these are listed separately although the observations were simultaneous.}
\begin{tabular}{lcccc}        
\label{tab:observations-gaia14aae}
Telescope & Date & Filter & Exp.     & Num. \\
+ instrument           &      &        & [s]          & eclipses \\
\hline 
% \textit{Gaia14aae}\\
WHT +      &    2015 Jan 14    &      \filu\filg\fili                  &    3.0   &     3\\
ULTRACAM                   &    2015 Jan 15    &      \filu\filg\filr                  &    3.0   &     4\\
                   &    2015 Jan 16    &      \filu\filg\filr                  &    3.0   &     5\\
                   &    2015 Jan 17    &      \filu\filg\filr                  &    3.0   &     3\\
                   &    2015 May 23    &      \filu\filg\filr                  &    2.5   &     6\\
                   &    2015 Jun 22    &      \filu\filg\filr                  &    3.0   &     4\\
\\
TNT +     &    2015 Feb 28    &      \filkg                           &    7.0   &     2\\
ULTRASPEC                   &    2016 Mar 12    &      \filkg                           &    8.0   &     1\\
                   &    2016 Mar 13    &      \filkg                           &    5.0   &     2\\
                   &    2016 Mar 14    &      \filkg                           &    5.0   &     3\\
                   &    2016 Mar 15    &      \filg                            &    5.0   &     2\\
                   &    2017 Feb 21    &      \filkg                           &    5.0   &     1\\
                   &    2018 Feb 07    &      \filkg                           &    5.0   &     1\\
                   &    2018 Feb 09    &      \filkg                           &    5.0   &     4\\
                   &    2018 Feb 10    &      \filkg                           &    5.0   &     1\\
                   &    2018 Feb 11    &      \filkg                           &    5.0   &     2\\
                   &    2018 Feb 12    &      \filkg                           &    5.0   &     3\\
                   &    2020 Jan 26    &      \filkg                           &    6.0   &     2\\
                   &    2020 Feb 01    &      \filkg                           &    6.0   &     2\\
                   &    2020 Feb 02    &      \filkg                           &    6.0   &     1\\
                   &    2020 Feb 03    &      \filkg                           &    6.0   &     1\\
                   &    2020 Feb 23    &      \filkg                           &    6.0   &     3\\
                   &    2020 Mar 23    &      \filkg                           &    12.0  &     1\\
                   &    2021 Feb 01    &      \filkg                           &    8.3  &     1\\
                   
& 2023 Jan 15 & \filkg & 7.0 & 2\\
& 2023 Feb 08 & \filkg & 10.0 & 1\\
& 2024 Jan 22 & \filkg & 10.0 &1\\
& 2024 Mar 11 & \filkg & 6.0 & 1\\
\\
Hale +      &    2016 Aug 06    &      \filg\filr                       &    4.0   &     7\\
CHIMERA                   &    2016 Aug 07    &      \filg\filr                       &    4.0   &     6\\
                   &    2016 Aug 08    &      \filg\filr                       &    4.0   &     6\\
                   &    2021 Mar 12    &      \filg\filr                       &    3.0   &     1\\
&    2021 Jun 11    &      \filg\filr    &    3.0   &     1\\
&    2021 Jul 05    &      \filg\filr    &    3.0   &     1\\
&    2022 Aug 23    &      \filg\filr    &    3.0   &     1\\
\\
GTC +      &    2019 Jun 05    &      \filgs\filrs\filis   &    3.0   &     4\\
HiPERCAM     &        &      \filus\filzs   &    9.0   &     4\\
% & 2024 Mar 06 & \filus\filgs\filrs\filis\filzs & 2.7 & \\ % skipping as only got half an eclipse
  & 2024 Apr 16 & \filgs\filrs\filis & 2.7 & 1 \\
  &  & \filus\filzs & 8.1 & 1 \\
\hline
\end{tabular}
\end{table}

\section{Unbinned Ephemerides}

\begin{figure}
\includegraphics[width=\columnwidth]{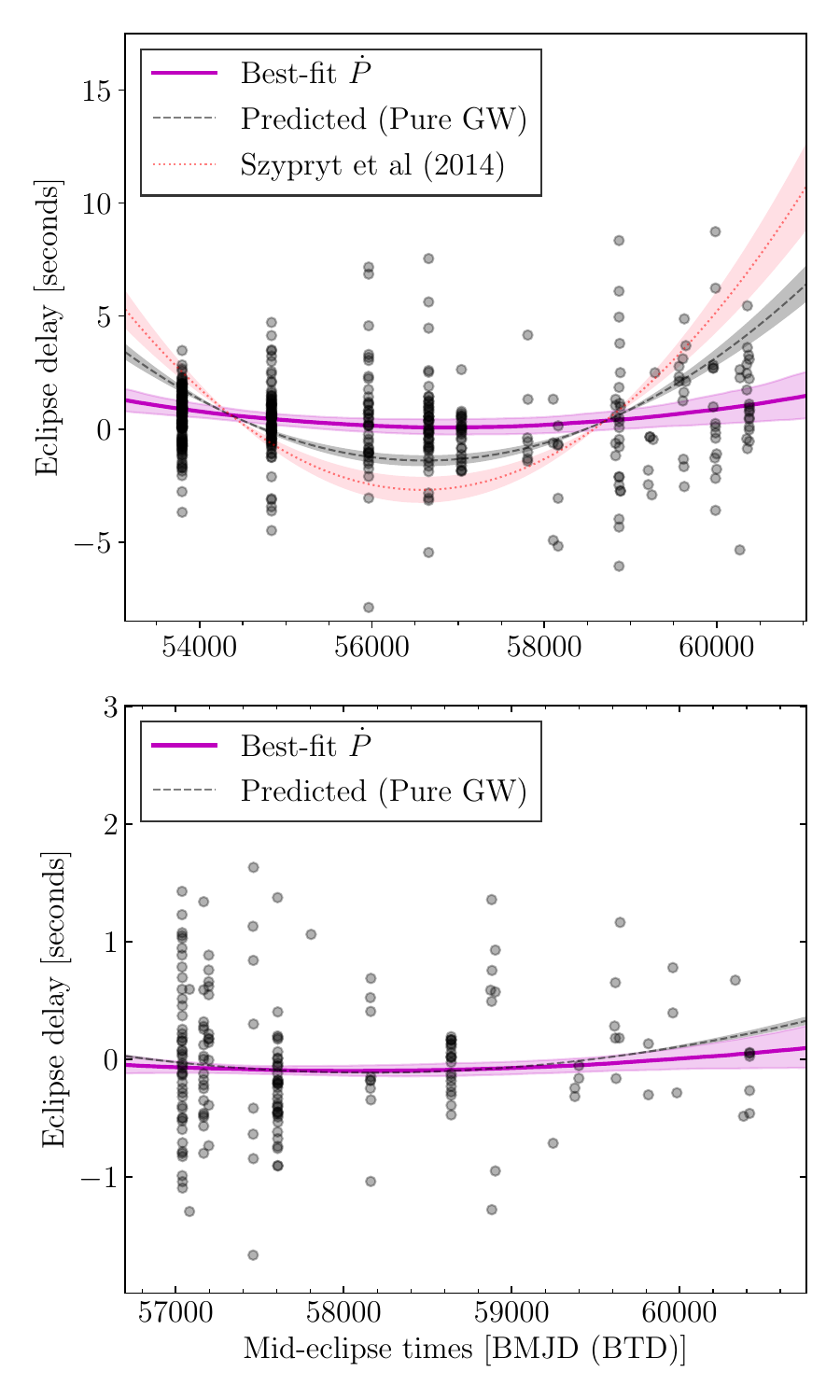}
\caption{Delay of the observed mid-eclipse times for YZ\,LMi (above) and Gaia14aae (below), compared to the expected mid-eclipse time under the assumption of a linear ephemeris. 
Note the change in axis scale compared to Fig.~\ref{fig:ephemeris}.
}
\label{fig:ephemeris-unbinned}
\end{figure}

In Fig.~\ref{fig:ephemeris-unbinned}, we show an unbinned version of Fig.~\ref{fig:ephemeris}.

%%%%%%%%%%%%%%%%%%%%%%%%%%%%%%%%%%%%%%%%%%%%%%%%%%

% Don't change these lines
\bsp	% typesetting comment
\label{lastpage}
\end{document}